\documentclass[9pt,useAMS,usenatbib]{mn2e}
\usepackage{color,graphicx,amssymb}

\newcommand{\apj}{ApJ}
\newcommand{\mnras}{MNRAS}

\title[Redshifted 21-cm Signals in the Dark Ages]{Redshifted 21-cm Signals in the Dark Ages}
\author[Kim and Pen]{Juhan Kim$^{1}$\thanks{E-mail:kjhan@cita.utoronto.ca}
and Ue-Li Pen$^{1}$\thanks{E-mail:pen@cita.utoronto.ca} \\
$^{1}$Canadian Institute for Theoretical Astrophysics, University of Toronto, 60 St. George Street,
Toronto, ON M5S 3H8, Canada
}

\begin{document}
\date{Accepted 20xx . Received 20xx; in original form 20xx}
\pagerange{\pageref{firstpage}--\pageref{lastpage}} \pubyear{2002}
\maketitle
\label{firstpage}

\begin{abstract}
We have carried out semianalytic simulations to build redshifted 
21-cm maps in the dark ages.  An entropy-floor model is adopted 
for planting protogalaxies in simulated minihaloes. 
The model allocates gas quantities such as baryonic mass and 
temperature to every $N$-body particle and extensively exploits 
the particle nature of the data in the subsequent analysis.
We have found that the number density of simulated minihaloes 
in the early universe is well described by the Sheth \& Tormen function
and consequently the signal powers of simulated minihaloes are far greater than 
the Press \& Schechter prediction presented by \citet{furlanetto06b}.
Even though the matter power spectrum measured in the halo particles 
at $z=15$ is about an order of magnitude smaller than the intergalactic 
medium (IGM), the 21-cm signal fluctuations of haloes are, to the contrary, 
one order of magnitude higher than the embedding adiabatic IGM 
on scales, $k\lesssim 10~h{\rm Mpc}^{-1}$.
But their spectral shapes are almost same to each other.
We have found that the adiabatic signal power on large scales lies
between the linear predictions of the infinite spin-temperature model
($T_s\gg T_{\rm cmb}$) and the model with the uniform spin temperature equal to 
background value ($T_s = T_s^{bg}$).
Higher preheating temperature (or higher background entropy)
makes the power spectrum of signals more flattened 
because the hotter IGM signals are more thermally broadened and
minihalo fluctuations dominating on small scales are more severely suppressed 
by the higher background entropy.  
Therefore, this model-dependent power spectrum slope measured 
on the scale of $100\le k \le 1000 ~h{\rm Mpc}^{-1}$ will enable us to easily 
determine a best-matching halo $+$ IGM model in future observations.

\end{abstract}


\section{Introduction}
%The neutral hydrogen emits or absorbs photons at wavelength $\lambda_{21}=21.1$cm.
%This happens when the electron at the state of 1$S$ 
%transits between two energy levels of spin alignments with the necleus. 
%Because the basic physics of radiative transfer is well known,

Until recently, the Large-scale Structures (LSS) of the universe 
and the Cosmic Microwave Background (CMB) radiations have been 
two principal research areas in the astrophysical cosmology.
They are distinct from each other in their nonlinearity 
and observation scales.
The LSS \citep{jarrett04,gott05,peebles80} is a complex nonlinear 
structure forming relatively recently (compared to the CMB)
while the CMB radiation \citep{penzias65,smoot93,spergel03}
comes through the last scattering surface when the universe is 
still in the linear regime just 0.4 million years after the Big Bang. 
The LSS is a cumulative result of nonlinear gravitational evolution
over the entire age of the universe ($t_{\rm age}\sim$ 13.7 billion years) 
and, on the other hand, the CMB glow is a transient event when 
lights and baryons are decoupled from each other as the universe 
cools down below $T \sim 3000$K (we simply ignore the subsequent 
interactions of CMB photons with cosmic objects such as evolving 
gravitational potential and free electrons when they travel from 
the CMB photosphere to us).
%The CMB-related physics are ``relatively'' simpler and more straightforward 
%than the LSS physics.

However, between these two epochs, there is another era ($11 \lesssim z \lesssim 1000$)
called the dark ages that have recently been regarded as a big cosmological 
reservoir of rich information on the early universe \citep{loeb04,pen04,furlanetto09a,bowman09}
and could help us pin down cosmic parameters
more accurately \citep{cooray08,mao08,furlanetto09b}.
This epoch is essential to the study of the cosmology
since it can cover the shortcomings of the CMB and LSS studies.
Among those shortcomings are the difficulty of CMB observation on galactic scales
for a more accurate determination of the power spectral index of matter field
and the difficulties arising during the interpretation of LSS observations
due to the nonlinear evolution and complex biasing effects.

In the dark ages, the density fluctuation is still in linear or 
quasi-linear regimes
and the universe remains dark by large because there seldom exist 
strong photon-emitting sources such as stars or active galactic neuclei (AGNs).
However, there is an astrophysical observable,
the neutral hydrogen, which may imprint its existence
on the blackbody spectrum of cosmic background emission.
The neutral hydrogen emits or absorbs a photon at 21-cm 
as a bound electron flip-flopps its spin direction at the 1$S$ state.
As the basic physics of radiative transfer is well known,
we can easily decode the observed spectrum to get information on 
line-of-sight distribution of the hydrogen quantities
such as temperature, projected surface density, and line-of-sight velocity.
Observations of the abundant neutral hydrogen in the dark ages
would, consequently, provide us with a powerful window of opportunity 
for obtaining wealthy physics on the IGM and birth of protogalaxies
at the dawn of the universe.

%The cosmic history of the 21-cm emission and absorption is a result of
%the competition between the photon and neutral hydrogen:
The emission and absorption of the redshifted 21-cm line on the background CMB spectrum
indicate the interactions between the photon and neutral hydrogen:
the spin temperature of a neutral hydrogen is governed by 
the CMB temperature, color temperature of $Ly\alpha$ photon, and the 
kinetic temperature of the hydrogen gas \citep{purcell56,field59,furlanetto06a,hirata06}. 
Even after the decoupling epoch, the baryonic gas temperature
is tightly coupled to the CMB temperature until $z\sim 300$ \citep{furlanetto06a}
after which the gas temperature begins to drop more rapidly ($T_g \sim (1+z)^2$)
than the CMB temperature ($T_{\rm cmb} \sim (1+z)$).
Because the gas density is still sufficiently high, the spin temperature of gas 
is coupled to the gas temperature through atom-atom collisions until $z\sim 100$.
However, as the gas density drops, $\rho_g\sim (1+z)^3$, 
the hydrogen density is not any more sufficient to hold the spin temperature.
As a result, after $z\sim25$ the CMB radiation has played a dominant role 
in driving the spin temperature.

There are several ongoing and forthcoming
projects for observing the redshifted signals 
with the state-of-the-art interferometry techniques.
The Giant Metrewave Radio Telescope\footnote{http://www.gmrt.ncra.tifr.res.in} 
(GMRT, \citealt{pen08}) has been set up in India 
with 30 parabolic dishes of 45m diameter spanning 25km in a Y-shaped configuration
for the target wavelength from 50 to $1,420$MHz.
One of main observational targets is the emission from neutral hydrogen
in the protogalaxies or protoclusters between redshifts 3 and 10.
The Murchison Widefield Array\footnote{http://www.mwatelescope.org} 
(MWA, \citealt{lidz08}) is designed for
the observation of the amplitude and slope of redshifted 21-cm power spectrum on scales,
$k \sim 0.1$ -- $1~ h{\rm Mpc^{-1}}$, especially for the reionization epoch.
The LOFAR\footnote{http://www.lofar.org} (Low Frequency ARray, \citealt{zaroubi05}) 
will use the arrays of dipole antennas built across the European countries 
for the observations of redshifted hydrogen signals in the epoch of reionization.
The PrimevAl Structure Telescope\footnote{http://web.phys.cmu.edu/~past/}
(PaST, \citealt{peterson04}) will consist
of log-periodic antennas in China targeting the first luminous objects 
in the epoch of reionization.
And the Square Kilometer Array\footnote{http://www.skatelescope.org/} (SKA)
will be built in the southern hemisphere
to map three-dimensional distributions of neutral hydrogen in the dark ages
and reionization eras.

The purpose of this paper is two-folded.  
Firstly, we want to apply an entropy-floor model to semianalytically 
simulating protogalaxies in minihaloes \citep{iliev02,furlanetto02,
martel03,shapiro06} from $N$-body particles.
To exploit the particle nature of the simulation data, we implement a new method 
to accurately measure the optical depth even in highly dense regions.
Secondly, by applying the power spectrum analysis to the generated maps 
we want to fully assess the minihalo contribution to the diffuse backgrounds.
However, it is a much challenging job to observe individual minihaloes 
at the low frequency of redshifted 21-cm photons even in the recent future
due to the small angular size ($\theta_{\rm halo} \lesssim$ a few arcseconds)
and the weak signals ($T_b \lesssim$ a few tens mK). 
Most of the ongoing projects have angular resolutions one or two orders 
poorer than needed to detect individual haloes.
But the signals of minihaloes over the diffuse background 
can be detectable by the current planned interferometers
and, moreover, the rapid evolving radio astronomy 
will make it possible to push back the current resolution limit 
beyond the minihalo scales in the future.
Therefore, it is worthwhile to study the minihalo signals in the cosmic context.

For this analysis, we run cosmic $N$-body simulations and apply the entropy-floor
model \citep{pen99,kaiser91,voit05,ostriker05} 
to build baryonic contents (or protogalaxies) in virialized dark minihaloes. 
Also the peculiar velocity and thermal broadening 
are included in the method to simulate the observed redshift distortions.
The equation of absorption along the line of sight is fully solved even 
in halo regions where neutral hydrogen is so dense that 
the exact measurement of optical depth is much more important than anything else
for the bright sources.

We perform a dark-age benchmark test for redshifted signals of neutral hydrogen
at $z=15$ when the most of gas still remains neutral.
In this paper, we do not include the effects of $Ly\alpha$ and ionizing photons 
which are crucial for the study of the reionziation era.
The onset of reionization epoch is still unclear because it is still 
beyond current observational barriers.
However, we are able to get a clue from observations of quasars and CMB.
The presence of the Gunn-Peterson trough \citep{gunn65} in the quasar spectrum 
implies that there are abundant neutral hydrogens in the IGM beyond 
$z\sim 6$ \citep{becker01} which implies that the reionization process 
is completed after this redshift.
Also from the WMAP observations, the angular power spectrum of CMB
consistently favors the extended ionization process \citep{dunkley09}
that the reionization started at $z\sim 11$ 
and finished at $z=7$ \citep{spergel07}.
Therefore, the selection of $z=15$ is adequate for the benchmark test 
for the dark ages.

The content of the paper is as follows:
the basic physics related to the 21-cm emission and absoption 
are given in Section \ref{physics}.
%Section 3 is devoted to the descriptions of the minihalo characteristics including
%the ionization fraction, profiles of density, spin temperature, and the brightness temperatures.
We introduce a new and robust method to generate the redshifted 21-cm signal maps
from the $N$-body simulation particles in Section \ref{minihaloes}. 
In Section \ref{application} we describe how to measure the optical depth 
and how to achieve the doppler and thermal broadenings with the $N$-body particles.
Section \ref{simulation} briefly describes the simulation and halo findings.
Section \ref{map} presents the resulting maps of various models and the effect 
of the thermal broadening and Doppler shift on the image of minihaloes.  
Also the power spectrum analysis is given in the latter part of this section.
We conclude with several arguements and remarks in Section \ref{conclusion}.
Appendix \ref{app:powerspectrum} is devoted to provide a quick look at the
differences of the linear power spectrum between various methods.
We extensively show how to determine the initial redshift of the simulation
in Appendix \ref{app:zeldovich}.

In this paper, we assume a concordance $\rm \Lambda$CDM (cold dark matter) cosmology 
consistent with the WMAP 5-year data \citep{komatsu09}.
We set the current CMB temperature to be $T_{\rm cmb}(0) = 2.725$,
and the helium mass fraction, $Y_p=0.24$ \citep{schramm98}.
We also assume that there is no $Ly\alpha$ photons so that the
Wouthuysen-Field effect \citep{baek08,hirata06} is fully neglected.
The frequency of the 21-cm line in a rest frame is $\nu_{21}=1.42$ GHz.
We mix the use of terms, haloes and minihaloes, for the same sense. 
Also baryonic matter means the gas mixture of hydrogen and helium.
Gas temperature generally implies gas kinetic temperature.

\section[]{21-cm Emission and Absorption Lines}
\label{physics}
The spin temperature ($T_s$) is a weighted sum of three competing temperatures 
such as the CMB ($T_{\rm cmb}$), color ($T_\alpha$), and gas kinetic 
temperatures ($T_g$) \citep{purcell56,furlanetto06a,shapiro06}:
\begin{equation}
\label{Ts}
T_s = {T_{\rm cmb} + y_\alpha T_\alpha + y_c T_g \over 1+y_\alpha + y_c},
\end{equation}
where the color temperature is related to the $Ly\alpha$ photons.
The collisional weighting coefficient is scaled to the CMB contribution as
\begin{equation}
y_c = {T_\star C_{10} \over T_g A_{10}},
\end{equation}
where $C_{10}$ is the collisional de-excitattion rate \citep{purcell56} and $A_{10}$ is the 
Einstein spontaneous emission coefficient.
We set $y_\alpha = 0$ assuming that there are no stars and AGNs which
could emit the $Ly\alpha$ photons and ionize neutral atoms in the medium.

\begin{figure}
\center
\includegraphics[scale=0.45]{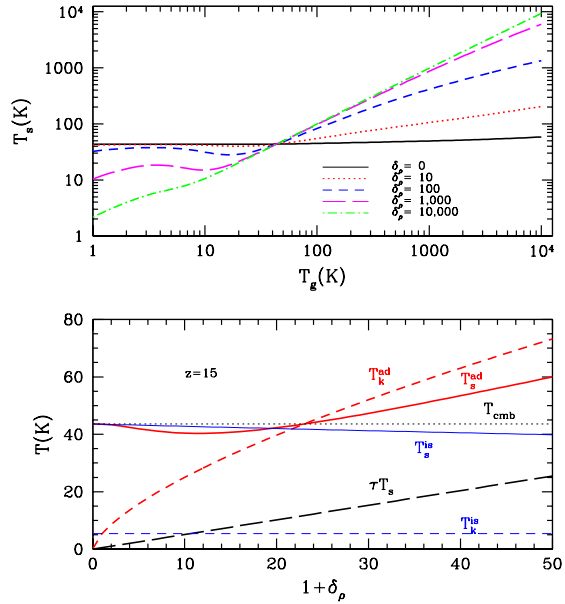}
\caption{
{\it Lower}: Kinetic ({\it short-dashed}) and spin ({\it solid line}) temperature distributions 
for the adabatic ({\it thick}) and isothermal ({\it thin line}) IGM models at $z=15$.
The CMB temperature is marked by the dotted curve.
Also we overplot the product of the spin temperature and optical depth ($\tau T_s$)
in a long-dashed line. This quantity does not depend on the temperature model of the medium.
{\it Upper}: Distribution of spin temperature as a function of kinetic temperature for
various density contrast at $z=15$.
}\label{fig-spin}
\end{figure}
The lower panel of Figure \ref{fig-spin} shows distributions of spin and 
kinetic temperatures of the IGM as a function of the local density contrast at $z=15$.
Here the gas temperature is assumed to follow the adiabatic ({\it thick}) 
or isothermal ({\it thin}) processes.
The adiabatic spin temperature moves toward the kinetic temperature as the density  
contrast grows and the isothermal spin temperature is dropping to the 
cold gas temperature which is below the CMB temperature at this redshift.
It is interesting to note that the overdense region could be colder than the CMB,
provided $\delta \le 21$.
The spin temperature is not tightly coupled to the kinetic temperature even 
at high-density contrasts because the absolute density is not 
sufficiently high at $z=15$.
The upper panel shows the distribution of the spin temperature as a function of the
kinetic temperature for the gas density contrasts, $\delta_{\rho}=0$, 10, 100, 1000, and 10000.
In the mean field ($\delta_\rho = 0$), the spin temperature is nearly invariant of
the kinetic temperature up to $T_g = 10^4$K because the gas density is 
much low. 
If the density contrast rises up to 1000, the spin temperature is almost 
coupled to the kinetic temperature.
Therefore, we expect that the spin temperature in the virialized 
haloes may strongly be coupled with the gas temperature because of the high density.
On the other hand, in the mean field the spin temperature is nearly invariant 
of the gas temperature.

One can calculate the brightness temperature over the background 
CMB temperature ($T_{\rm cmb}$) as
\begin{equation}
\Delta T_b (\nu)=
\int_{0}^{\tau(\nu)}
{T_s(z^\prime)-T_{\rm cmb} (z^\prime) \over {1+z^\prime}}
e^{-\tau^\prime(\nu)} {d \tau^\prime(\nu)},
\label{Tb}
\end{equation}
where $\Delta T_b \equiv T_b - T_{\rm cmb}$, $z^\prime$ is 
the redshift of the source, $T_s$ is its spin temperature,
and $\tau^\prime(\nu)$ is the optical depth to the source at 
the frequency, $\nu$.
The increase of optical depth due to a gas element at $z^\prime$ 
may be computed as \citep{shapiro06},
\begin{equation}
\label{tau}
%{d \tau^\prime (\nu)} = {3\lambda_0^3A_{10} \over 32\pi}
{d \tau(\nu)} = {3\lambda_0^3A_{10} \over 32\pi}
{T_\star\over T_s(z^\prime)} {n_{\rm HI}(z^\prime)\over H(z^\prime)} 
\phi(\nu^\prime-\nu) d\nu^\prime,
\end{equation}
where we have used the relations, $d\ln(\nu^\prime) = d\ln(1+z^\prime)$ and
$n_{\rm HI} = (1-\chi_{\rm HII}) \rho_{g}  (1-Y_p)/m_{\rm H}$.
The gas density is related to background gas density as $\rho_g = (1+\delta_g) \varrho_g^{bg}$ where
$\varrho_g^{bg} = \rho_c(z) \Omega_b$ and $\rho_c(z)$ is the critical density at $z$.
At $z=15$, the mean ionization fraction of IGM hydrogen is 
$\chi_{\rm HII}^{\rm IGM}=1.92\times10^{-4}$ which is obtained from the RECFAST package 
\footnote{http://www.astro.ubc.ca/people/scott/recfast.html} \citep{seager99}.
%We will fix the IGM ionization fraction to this value in this study
%Hereafter, we assume that the IGM is uniformly ionized at this ratio value 
%irrecpective of the background temperature and density fluctuations
%while we compute ionization fractions in the halo regions by solving the Saha equation,
%which will be discussed in the following section.
Hereafter, we uniformly apply this RECFAST value to all hydrogens 
even though they are located in halo regions.
This assumption is correct
because the virialized temperature and the gas density of a minihalo
($M_h \lesssim 10^7 h^{-1}{\rm M_\odot}$) is not as high as to ionize the 
hydrogen atom through a collisional process.
The ionization time scale is nearly infinite.

The Gaussian velocity dispersion spreads the profile, $\phi(\nu)$, 
of emission or absorption lines along the line of sight as,
\begin{equation}
\phi(\nu^\prime-\nu) = {1\over \sqrt{2\pi} \sigma(\nu^\prime) } e^{-{1\over 2}
\left({\nu^\prime -\nu}\right)^2 / \sigma^2(\nu^\prime)}
\end{equation}
where the observed (one-dimensional) frequency dispersion $\sigma(\nu^\prime)$ is related to the
kinetic temperature of gas as 
$\sigma(\nu^\prime) = (\nu^\prime/c)(kT_g/m_{\rm H})^{1/2}$.
Here, $m_{\rm H}$ is the hydrogen mass.
The combination of the cosmic redshift and Doppler shift in the frequency 
is expressed as 
\begin{equation}
\nu^\prime = {\nu_{21}\over 1+z^\prime} {1+(v/c)\over \sqrt{1-(v/c)^2}},
\label{dop}
\end{equation}
where $v$ is the peculiar velocity toward the observer.
In the mean field of no peculiar velocity, we can get 
$\tau(0) = ({3\lambda_0^3/ 32\pi) (A_{10} n_{\rm HI}(z)/ H(z)}) ({T_\star/T_s(z)})$
by assuming that $\phi(\nu) = \delta (\nu)$ where $\delta(\nu)$ is the delta function.
In the limit of $\tau \ll 1$, equation (\ref{Tb}) can be approximated to
\begin{eqnarray}
\nonumber
\Delta T_b &\simeq& 34 \chi_{\rm HI} (1+\delta) \left({\Omega_b h^2\over 0.023}\right)
\left({0.15 \over \Omega_m h^2 }{1+z\over 16}\right)^{1/2} \\
&&\left({T_s-T_{\rm cmb}(z)\over T_s}\right) {\rm mK},
\end{eqnarray}
where we applied the approximation that $H(z) \simeq H_0 \Omega_m^{0.5} (1+z)^{1.5}$.
In Figure \ref{fig-spin}, we show 
the distribution of $\tau T_s$
which is equal to the unredshifted brightness temperature in the limit of $\tau \ll 1$.

\section[]{Minihaloes}
\label{minihaloes}

\subsection{Isothermal Models}
In the singular isothermal sphere (SIS) model, 
the baryonic density of a virialized halo follows a simple power law as
$\rho_g(r) = v_c^2\Omega_b/4\pi G \Omega_m r^2$, where the circular velocity is
defined as $v_c^2 \equiv GM_v/R_v$. The real-space virial radius ($R_v$)
is defined by the extent to which the mean density of 
the halo is $v_{178}\rho_c(z)$. 
%The virial overdensity can be obtained 
%from the fitting function of $v_{178}= 18\pi^2 + 82y-39y^2$ with 
%$y\equiv \Omega_m(z)-1$ \citep{bryan98}.
At high redshift such as $z=15$, it is sufficient to set $v_{178}\simeq 178$
\citep{bryan98}.
We can derive the isothermal kinetic temperature of the baryonic sphere
using the energy relation, 
\begin{equation}
{3\over2}\left({k_B T_v\over \mu_{\rm H} m_{\rm H}}\right) = {v_c^2\over 2},
\label{virialT}
\end{equation}
where $\mu_{\rm H}$ is the mean molecular weight of a gas mixture.
The mean molecular weight is computed by the helium mass fraction, $Y_p$,
and the ionization of hydrogen atom, $\chi_{\rm HII}$.

However, the SIS model breaks down in the central region 
where the second law of thermodynamics might be violated: 
the central entropy happens to be less than the background IGM entropy.
According to the second law of thermodynamics, the entropy of a system
should only increase and, therefore, haloes forming out of the IGM
should have an entropy value equal to or larger than the IGM entropy.
The astrophysical entropy density is defined by $ K \equiv {T_g \rho_g^{-2/3}}$,
which is different from the classical definition\footnote{the standard 
definition of entropy is related to the astrophysical conventional form 
as $s \propto \ln K^{3/2}$} but widely used because of its compact form 
beneficial for the analysis of the inner structures of observed clusters \citep{voit05,mitchell08}.
During the adiabatic cosmic expansion, the background entropy 
($K_{\rm IGM}$) is fixed with time because $T_g\propto(1+z)^2$ 
and $\rho_g \propto(1+z)^3$.
However, the halo entropy in the SIS model is a rising function with radius, 
$K_{\rm SIS} \propto r^{(4/3)}$, and, as a result, the entropy 
below a critical radius may happen be less than $K_{\rm IGM}$.

There have been many observational evidences \citep{voit05,balogh06}
for the entropy floor in the inner region of cluster haloes and
many researchers have proposed various models for describing the flat core entropy
\citep{pen99,oh03,xu03,roychowdhury04,ostriker05}.
In this work, we adopt the entropy-floor semi-isothermal sphere (EIS) model
of the density distribution given by \citet{pen99} who proposed a density
profile as
\begin{eqnarray}
\label{eisrho}
\rho_g(r) = \left({ v_c^2\Omega_b\over4\pi G\Omega_m}\right)
\left\{ \begin{array}{ll}
{1\over R_c^2}\left[ 1 - {12\over25} \log \left({r\over R_c}\right)\right]^{3/2} 
&\textrm{if } r<R_c , \\ \\
\left({1\over r^2}\right) &\textrm{otherwise,}
\end{array}\right.
\label{ueli}
\end{eqnarray}
where $R_c$ is the core size.

The temperature outside the core is assumed to be isothermal
and simply measured by equation (\ref{virialT}).
And the core radius can be derived 
by equalising the entropy at the core boundary ($r=R_c$)
to the background entropy, $K_{\rm IGM}$, as
%\begin{eqnarray}
%\nonumber
$T_v\rho_g^{-2/3}(R_c) = T_g^{bg}(z) {{\varrho}_g^{bg}}^{-2/3}(z)$,
%\\
%&=& T_{bg}(0) {\varrho}_{bg}^{-2/3}(0),
%\end{eqnarray}
where $T_g^{bg}(z)$ is the background gas temperature at $z$.
If combined with equations (\ref{virialT}) and (\ref{ueli}),
this relation leads to
\begin{eqnarray}
\nonumber
\left({R_c\over R_v}\right) 
&=& 
\sqrt{3\over2}
\left({k_B T_g(0) (1+z)\over \mu_{\rm H} m_{\rm H} G}\right)^{3/4}
\left({v_{178}\over \pi \rho_c(0) }\right)^{1/4}
\sqrt{{1\over\Omega_m M_v}}\\
\nonumber
&=& 1.808
\left({1+z \over 16 }\right)^{3/4} \left({\mu_{\rm H} \over 1.219}\right)^{-3/4}
\left({\Omega_m\over 0.258 }\right)^{-1/2}\\
&&
\left({T_g^{bg}(0)\over 0.0214K} \right)^{3/4}
\left({M_v\over 10^4~h^{-1}{\rm M_\odot}}\right)^{-1/2}
\left({v_{178}\over 178}\right)^{1/4}.
\end{eqnarray}
We note that the relative core size is 
anti-correlated with the halo mass indicating that smaller minihaloes
are more strongly affected by the entropy constraint.

%There is another relation for a halo quantity: 
%the mass ratio between the baryon and the matter contents in a halo.
Also we can derive another relation between the halo mass and the 
infall baryonic fraction.
Here, the infall baryonic fraction is defined by the 
mass ratio of infall gas to the entire gas which is initially located
in the collapsed ``Lagrangian volume'' of the dark matter.
Since the total gas mass in a halo is computed 
with $M_g = 4\pi\int_0^{R_v}\rho_g(r) r^2 dr$,
the infall mass fraction of gas is simply given by
\begin{displaymath}
f_g \equiv
\left({M_g\over M_v}\right)\left({\Omega_b \over\Omega_m }\right)^{-1}= \left\{ \begin{array}{ll}
1 - 0.581 \left({R_c\over R_v}\right) &\textrm{if } R_c < R_v,  \\
A\left[\left({R_c\over R_v}\right)^{-1}\right] &\textrm{otherwise},
\end{array}\right.
\end{displaymath}
where 
\begin{eqnarray}
\nonumber
A(a) &\equiv& {2\over125}e^{25\over4}{\sqrt{\pi}\over a} 
\left[{1-\textrm{erf}\left({1\over2}\sqrt{25-12\ln a}\right)}\right]\\
&+&{a^2\over375}\left(31-12\ln a\right)\sqrt{25-12\ln a}.
\end{eqnarray}
If $f_g=1$, all baryonic matter settles down to the halo centre. 
And $f_g=0$ means that there has been no baryonic collapse 
and, consequently, the halo has no galaxy or no gas in it.
The gas-temperature profile of a halo is simply measured as
\begin{displaymath}
T_g(r) =  \left\{ \begin{array}{ll}
T_{g} (0) \left({\rho_g(r) \over \rho_c(0)\Omega_b}\right)^{2/3} &\textrm{if } r<R_c, \\ \\
T_v &\textrm{otherwise}.
\end{array}\right.
\end{displaymath}

\begin{figure}
\center
\includegraphics[scale=0.5]{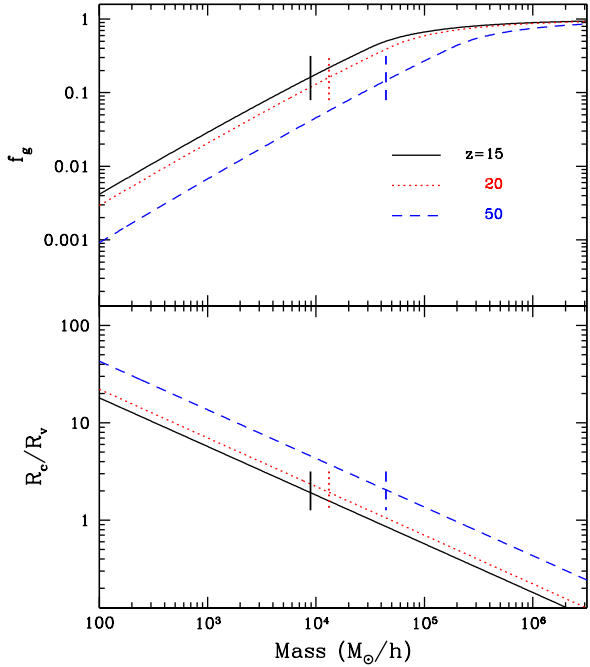}
%\plotone[scale=0.5]{pen.eps}
\caption{
Mass dependence of relative core size, $R_c/R_v$ ({\it bottom}), and 
infall baryonic fraction, $f_g$ ({\it top panel}), at $z=15$, 20, and 50.
Vertical bars mark the Jean's mass at those epochs.
}\label{fig-pen}
\end{figure}

Figure \ref{fig-pen} shows the dependences of the relative 
core size ({\it bottom}) and infall baryonic fraction 
({\it top}) on the halo mass.
The core size is an increasing function of redshift at a fixed halo mass
and, consequently, the infall fraction $f_g$ is decreasing with redshift.
We may check whether the entropy-floor model contains the Jean's mass condition
\citep{gnedin98} which is defined as the mininum mass 
of a spherical overdense region that can gravitationally collapse overcoming
the resistant thermal pressure.
In the figure, we note
that the Jean's mass scale (calculated by Eq. 23 of \citealt{shapiro06})
is roughly corresponding to the mass of a halo 
whose fractional infall-gas mass is $f_g \sim 0.1$ -- 0.2 
(or the corresponding relative core size is $(R_c/R_v)\sim$ 2 -- 3).
This indicates that, on the Jean's scale, 
80 to 90\% of total baryonic mass inside the halo Lagrangian 
volume may not infall to the halo centre.
So it is reasonable to think that the entropy-floor model
may inherently have the Jean's mass criterion.
Therefore, we may simply skip the setting of the minimum halo mass
usually applied to the semianalytic process to build protogalaxies in minihaloes.

\subsection{Spin \& Brightness Temperature Profile}
\begin{figure}
\center
\includegraphics[scale=0.5]{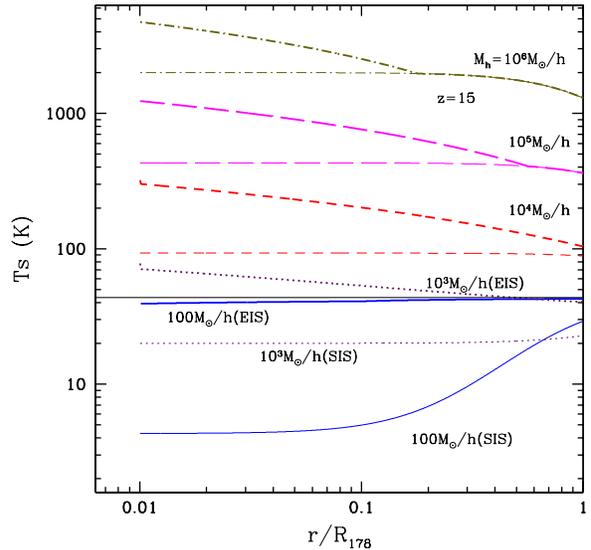}
\caption{ 
The spin temperature profile for virialized minihaloes in the EIS ({\it thick})
and SIS ({\it thin}) model at $z=15$. 
For clarity, we add the model name in parenthesis
for haloes of mass $M_h=100$ and $10^3 h^{-1}{\rm M_\odot}$.
The thin solid horizontal line marks the CMB temperature at the epoch.
}\label{fig-hts}
\end{figure}

We investigate the distributions of spin temperature in the SIS and 
EIS halo models.  The spin temperature is derived by equation 
(\ref{Ts}) after we measure the gas density and temperature in each 
halo model.  Figure \ref{fig-hts} shows radial distributions of spin 
temperatures for various halo masses.  The SIS halo ({\it thin}) has 
a flat spin distribution in the inner region.  And haloes of mass 
below $M_s \sim 3\times10^3~h^{-1}{\rm M_\odot}$ have a spin 
temperature lower than the background CMB temperature. In the EIS 
model ({\it thick}) the characteristic mass scale where the spin 
temperature is same as the CMB temperature, decreases down to 
$M_s \sim 10^2~h^{-1}{\rm M_\odot}$.  From this figure, we know 
that $T_s$ is approaching $T_{\rm cmb}$ at the outer boundary of 
a halo mainly due to the low gas density: $T_s$ is decoupled from 
$T_g$ and coupled to $T_{\rm cmb}$ through the Compton scattering.
In the inner part of the SIS halo the spin temperature is saturated 
to the uniform gas temperature.  However, the spin temperature in 
a core of the EIS halo keeps rising for $M_h \ge 10^3 {\rm M_\odot}$ 
simply because the gas temperature increases with radius in the core.
%An intriguing feature is observed in the EIS haloes of $M = 10^6~h^{-1}{\rm M_\odot}$ 
%which shows a sharp drop of the spin temperature in the core.
%This is because the core is hot enough to ionize most hydrogen atoms and
%this ionization reduces the collisional excitation rate finally 
%leading to lowering the spin temperature.
%In most cases, collisions between the neutral hydrogens and ionized atoms are
%generally subdominant \citep{furlanetto06a} and can be neglected in calculating $C_{10}$.

\begin{figure}
\center
\includegraphics[scale=0.5]{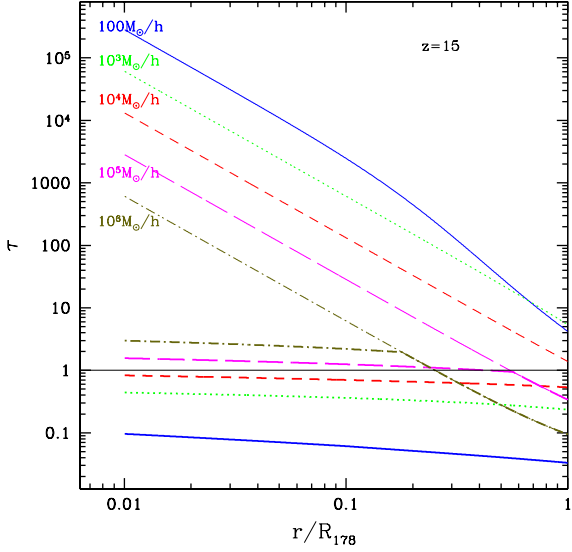}
\caption{
Profiles of the optical depth in EIS ({\it thick}) and SIS 
({\it thin curves}) halo models at $z=15$.  Halo masses 
are written besides the curves of the SIS model.
}\label{fig-tau}
\end{figure}

Integrating equation (\ref{tau}) over $\nu$ with an assumption 
that $\phi(\nu) = \delta(\nu)$ leads to the radial profile of 
the optical depth as 
\begin{eqnarray}
\nonumber
\tau(\nu,r) &\simeq& 0.51 \left({1+\delta_g(r)}\right) 
{\chi_{\rm HI}}
\left({T_s\over 1{\rm K}} \right)^{-1}
\left({\Omega_b h^2\over0.023}\right)\\
&&
\left({1+z\over 16}\right)^{3/2} 
\left({\Omega_mh^2\over 0.15}\right)^{-1/2}.
\label{eq-tau}
\end{eqnarray}
In Figure \ref{fig-tau} we show the dependence of optical depth 
both on the halo mass and on the radial distance.
In the SIS model ({\it thin}) the optical depth increases to the centre
but it declines with halo mass at a given relative radius (or at 
the same baryonic density) as noted by \citet{iliev02}.
Also it is interesting to note that beyond the radius of $0.2R_{178}$ 
the slope of the optical depth for the halo of $M=10^2~ h^{-1}{\rm M_\odot}$,
is steeper than more massive haloes.
This is because the radial distribution of $y_c$ (kinetic contribution 
to the spin temperature in Eq. (\ref{Ts})) becomes steeper due to 
the lower virial temperature.
Therefore, the spin temperature (also the brightness temperature)
approaches the CMB temperature more rapidly in this low-mass halo.
For more massive SIS haloes, the spin temperature is more tightly 
coupled to the isothermal gas temperature, so the profiles are 
parallel to each other: the distribution in the log-log scale shifts horizontally 
for different isothermal temperatures.  
The radial distribution of the optical depth follows that of the isothermal 
density as $\tau \propto (1+\delta_g(r))\propto r^{-2}$.
The optical depth in the core of an EIS halo ({\it thick}) 
is not so steep as in the SIS model because
the increase of gas density toward the halo centre
is partially offset by the increase of spin temperature (see Eq. \ref{eq-tau}).
%Due to the high ionization fraction in the center, 
%a halo of $M\gtrsim 10^6 ~h^{-1}{\rm M_\odot}$ has an optically 
%thin and transparent core.
%The inner transparent region is getting relatively bigger in size as the mass grows.
%And at last a halo of mass larger than $M_{\rm inv}=10^7 ~h^{-1}{\rm M_\odot}$ 
%becomes completely invisible to 21-cm observations after the outer region
%reaches the critical ionization temperature.
%But, even at this mass scale, the SIS halo still has a compact visible core 
%because of the retarded ionization for the high central density.

\begin{figure}
\center
\includegraphics[scale=0.5]{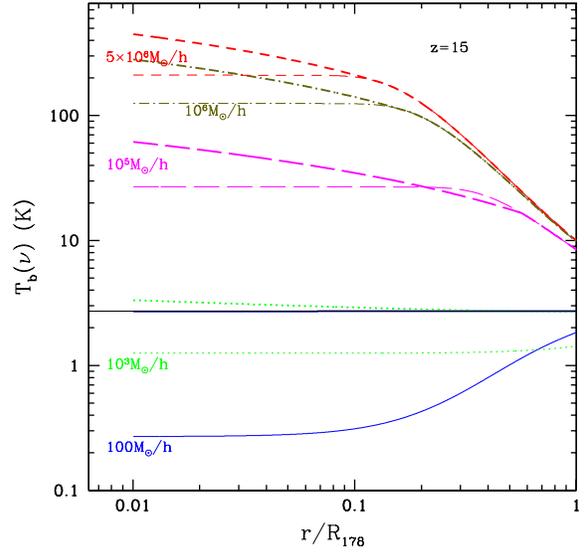}
\caption{
Profiles of brightness temperature of virialized minihaloes in the EIS 
({\it thick}) and SIS model ({\it thin curves}) at $z=15$. 
The horizontal line guides for the background CMB temperature.
The halo of $M = 100 ~h^{-1}{\rm M_\odot}$ in the EIS model,
has a temperature profile nearly overlapped with the background CMB temperature.
}\label{fig-htb}

\end{figure}

Because haloes are ususally optically thick as shown above,
it is important to apply a full expression of 
optical depth to the brightness temperature as,
\begin{equation}
\Delta T_b (\nu) = \left({T_s-T_{\rm cmb}(z) \over 1+z}\right)
\left( 1- e^{-\tau(\nu)}\right).
\label{eq-htb}
\end{equation}
It is valuable to note that
if $\tau \gg 1$, the brightness temperature is proportional to $(T_s-T_{\rm cmb})$.
On the other hand, if $\tau \ll 1$ and $T_s \gg T_{\rm cmb}$, 
we expect $\Delta T_b \propto (1+\delta_g)$.
Figure \ref{fig-htb} shows the predicted radial profile of the brightness temperature 
($T_b = \Delta T_b + T_{\rm cmb}(0)$) for various halo masses.
In the SIS model ({\it thin curves}), the temperature profile has 
two phases; the inner flat stage which is optically thick ($\tau \gg 1$),
and the outer power-law stage which is less thick ($\tau\lesssim$ a few). 
As expected, we observe that a massive halo with a high spin temperature
shows a power-law brightness temperature profile in the outer region because
the baryonic density is a most dominant factor in determining the brightness temperature.
Also in the EIS model ({\it thick}), the slopes of the brightness temperature 
and the spin temperature in the core region are almost same to each other
because of the nearly flat optical depth
which virtually fixs the second term of the right-hand side of equation (\ref{eq-htb}).
Minihaloes of $M=100 ~h^{-1} {\rm M_\odot}$ in the SIS model 
may be observed as cold spots but
those EIS counterparts can not be distinguished from the background
CMB temperature at $z=0$.
%Owing to the small optical thickness in the center of 
%the halo of $M=10^6~h^{-1}{\rm M_\odot}$,
%the brightness temperature drops rapidly to the background CMB temperature 
%and the halo seems to have a void core.

\subsection{Halo Contribution to Diffuse Backgrounds}
Now, we investigate the contribution of minihaloes to the 
diffuse backgrounds in terms of the brightness temperature 
and observed flux.
The averaged brightness temperature over the volume of a halo 
is defined in comoving space as
\begin{eqnarray}
\nonumber
\left<{\Delta T_b(M_v)}\right> &\equiv& 
{4\pi\int_0^{R_v^c} \Delta T_b(M_v,r,\theta,\phi) r^2dr 
\over V(M)}\\
&=& 3\int_0^{1} \Delta T(M_v,s) s^2ds,
\end{eqnarray}
where $V_h(M)$ is the comoving volume of a halo of mass $M$, $s\equiv r/R^c_v$,
and $R_v^c$ is the comoving virial radius.
If the sky is uniformly illuminated with a brightness temperature, ${\bar T_b}$,
the observed flux is 
${\bar T_b}\Delta\Omega_a\Delta\nu_{\rm obs}$
where $\Delta\Omega_a$ is the antenna beam solid angle and $\Delta\nu_{\rm obs}$ 
is the observation bandwith.
Therefore, the total contributon of minihaloes of mass $M$
to the diffuse background flux is a product of a single-halo contribution 
with the mean number density of haloes at the given mass as
\begin{eqnarray}
\label{meantb}
\nonumber
\bar{\Delta T_b}(M) \Delta\nu_{\rm obs} \Delta\Omega_{\rm a} &=&
\left<\Delta T_b (M_v)\right>\Delta\Omega_h \Delta\nu_{h}\\
&&
\left({dV\over dzd\Omega}\right)\Delta z \Delta\Omega_{\rm a} \Phi(M)
\end{eqnarray}
where $\Delta\Omega_h$ is the observed halo solid angle, 
$\Delta\nu_h$ is the effective halo size along the line of sight in frequency,
$\Phi(M)$ ($\equiv dn(M)/d\log_{10}M$)
is the number density of haloes of mass $M$,
$\Delta z=\Delta z(\Delta\nu_{\rm obs})$,
and $V$ is the survey comoving volume.
Using equation (\ref{meantb}), one may easily derive 
\begin{equation}
\bar{\Delta T_b}(M) = \left<\Delta T_b(M)\right> V_h(M) \Phi(M).
\end{equation}
Here, we have used ${dV/dzd\Omega} = {c d^2/H(z)}$ and 
$\Delta\Omega_h = {A/d^2},$ where $c$ is the speed of light,
$d$ is the comoving distance to the halo, and
$A$ is the geometrical cross section of the halo in the comoving space.
The product of last two terms is the total volume fraction
of minihaloes of mass $M$.
This result meets the reasonable expectation that 
the halo contribution to the brightness temperature should be 
a simple product of one-halo contribution 
with the comoving-volume fractions of the haloes of the same mass.
One may compare this equation with 
the one expressed in the real-space \citep{iliev02,shapiro06}.
The overall halo contribution to the diffuse backgrounds is measured by
$ \overline{\Delta T_b} = \int_0^\infty \bar{\Delta T_b}(M) d\log_{10} M$.
Also, the observed average flux from minihaloes can be measured by \citep{iliev02}
\begin{equation}
\delta \mathcal{F}_\nu(M) = {2k_B\over \lambda^2_{21}(z)}\bar{\Delta T_b}(M) \Delta\Omega_a,
\end{equation}
where we set $\Delta\Omega_a \equiv \pi(\Delta\theta_a/2)^2$
and $\Delta\theta_a$ is the simplified antenna beam angle.

\begin{figure}
\center
\includegraphics[scale=0.5]{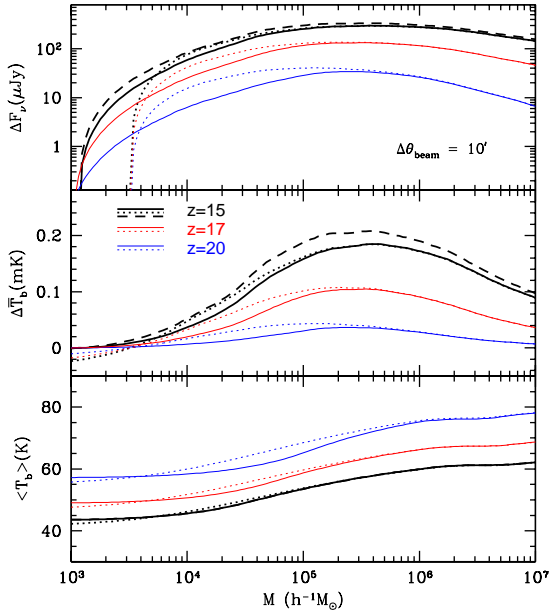}
\caption{ 
Brightness temperature of individual minihaloes ({\it bottom}), 
mean brightness temperature ({\it middle}),
and flux ({\it top}) observed with $\Delta\theta_a=10'$
as a function of halo mass $M$
at $z=15$ ({\it thick}), 17 ({\it intermediate}), and 20 ({\it thin lines}).
The solid and dotted lines are for the EIS and SIS models, respectively. 
The dashed curve show the EIS distribution computed by applying
the Eisenstein \& Hu power spectrum to the abundance of minihaloes at $z=15$.
}\label{fig-MTb}
\end{figure}

Figure \ref{fig-MTb} shows the distributions of three temperature-related observables
as a function of minihalo mass: the mean brightness temperature of a single minihalo
({\it bottom}), the mean brightness temperature in a unit solid angle
({\it middle}), and the mean minihalo flux 
({\it top panels}) received by an antenna of $\Delta\theta_a=10'$ at $z=15$, 17, and 20.
Each solid and dotted curves are for EIS and SIS models, respectively.
The contribution of minihaloes to the diffuse backgrounds
peaks around $M = 4\times 10^{5} ~h^{-1}{\rm M_\odot}$ 
and the peak position appears to be invariant of redshift.
In the SIS model,
there are negative contributions from minihaloes of mass 
$M\le 3\times10^3~h^{-1} {\rm M_\odot}$ while
the EIS haloes always make positive contribution to the diffuse backgrounds.
%Sharp drops are seen around  $M_{\rm inv}$, as expected,
%in the brightness temperature and flux distributions,
%and this characteristic mass scale decreases with redshift.
%The redshift dependence of the characteristic mass scale
%comes from the size dependence on the redshift at a given halo mass.
%At higher redshifts, the phyiscal size of same-mass haloes is smaller
%due to higher background density ($R_v \propto 1/(1+z)$) 
%and, consequently, the virial temperature is higher (see Eq. \ref{virialT})
%leading to higher ionization fraction in haloes.
The model dependences can be more easily seen on the lower-mass 
scales because lower-mass haloes have relatively bigger cores 
where the gas density and temperature more seriously deviate from the SIS models.
To determine  $\Phi(M)$ in this calculation, we have applied 
the power spectrum of the CAMB Source to the halo mass function
of Sheth \& Tormen (1999; ST).
Dashed curves show the resulting effect of the Eisenstein \& Hu 
(1998; hereafter EH) power spectrum on the distributions at $z=15$.
Appendix \ref{app:powerspectrum} specifies the differences between
these two power spectrum estimations.

\begin{figure}
\center
\includegraphics[scale=0.5]{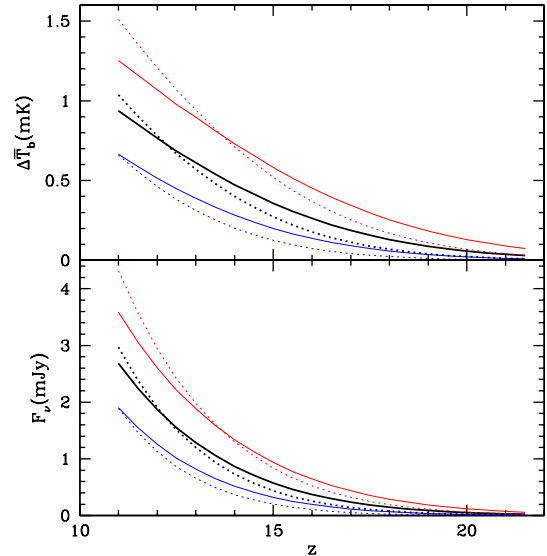}
\caption{
The minihalo emission. Solid lines are showing 
the background diffuse radiation from minihaloes predicted with the ST function
while dotted lines are based on the PS function.
Thick, intermediate, and thin lines are estimated with power indexes,
$n_s=0.96$, 1, and 0.92, respectively.
}\label{fig-zTb}
\end{figure}

As shown by \citet{iliev02}, we also check the redshift distribution 
of the brightness temperature and the corresponding flux emitted 
from our minihaloes.
In the upper panel of Figure \ref{fig-zTb}, the brightness temperature 
of minihaloes increases with time showing some significant deviations
among different spectral indexes of $n_s=0.96$ ({\it thick}), 
1 ({\it intermediate}), and 0.92 ({\it thin curves}).
Also the slopes of brightness temperature predicted from the 
Press \& Schechter function (PS; {\it dotted curves}) are steeper 
than the Seth \& Tormen.
This is because the PS function underestimates massive minihalo populations 
at high redshifts while overestimates the number density of less massive 
minihaloes at lower redshifts.
The minihalo flux over an antenna of $\Delta\theta_a=10'$ shows a similar distributions.

%where $q$ and $s$ are the ratios of the intermediat and minor radii with respect to 
%the major radius.
%From now on, we assume that $r$ is the radius in the major axis.
%To get $q$ and $s$ we measure the shape tensor, $\mathbf{T}$,
%from $N$ member particles,
%\begin{displaymath}
%\mathbf{T} = 
%\sum_i^N
%\left(
%\begin{array}{ccc}
%x_ix_i & x_iy_i & x_iz_i\\
%y_ix_i & y_iy_i & y_iz_i\\
%z_ix_i & z_iy_i & z_iz_i
%\end{array}
%\right),
%\end{displaymath}
%where $x$, $y$, and $z$ is the distance from the halo center of mass.
%Then, the eigenvalue ratios of $\mathbf{T}$ are $q$ and $s$ in decreasing order of
%magnitudes and eigenvectors are the directional cosines of the axes.
%In this case, the virial radius is switched to
%\begin{equation}
%r_{v} = \left({ 3M_{v}\over 712\pi qs \rho_c}\right)^{1/3}
%\end{equation}
\section[]{Applications to Simulations}
\label{application}
In this section, we introduce a new Lagrangian scheme 
for building brightness-temperature maps in the dark ages.
We show how to assign hydrogen gas density and gas temperature to each 
$N$-body particle. We, then, exploit the particle nature of the data
to compute the optical depth and to generate distortion maps 
adding the effects of the peculiar velocity and thermal broadening.

\subsection{Adiabatic Contraction in the IGM}
In the IGM, we have measured densities at a given position
using an adaptive smoothing kernel to enhance the spatial resolution.
The local density is measured with the 30 nearest neighbors 
by setting the smoothing length ($h_s$) equal to half the distance to the 30'th 
nearist neighbor.
Then, we estimate density at the positions of the IGM particles using
the smoothing kernel,
\begin{displaymath}
W_4(q) = \left\{ \begin{array}{ll}
\left(1-{3\over2} q^2 +{3\over4} q^3\right)/(\pi h_s^3) &\textrm{for }0<q\le1,\\
\left(2-q\right)^3/(4\pi h_s^3) &\textrm{for } 1<q\le2,\\
0 &\textrm{and otherwise},
\end{array}\right.
\end{displaymath}
where $q\equiv r/h_s$.
Under the assumptions of the adiabatic contraction
and no additional heating sources,
we can measure the kinetic temperature of the IGM gas using 
$
T_g = \left<T_g(z)\right> (1+\delta_g)^{\gamma-1}
$
where $\left<T_g(z)\right>$ is the kinetic temperature of mean backgrounds,
$\delta_g$ is the gas density contrast to the mean background,
and $\gamma=5/3$ for a  monoatomic ideal gas.

\subsection{Brightness Temperature of Simulated Particles}
The mean differential brightness temperature is discretized
according to the finite volume element of frequency range 
($\nu-\Delta\nu/2 \le {\bar \nu} \le \nu+\Delta\nu/2$) 
and cross section $\Delta S$ as,
\begin{eqnarray}
\nonumber
{\Delta T_b({\bar \nu})}&\equiv&
{1\over \Delta\nu}
\int_{\nu-\Delta \nu/2}^{\nu+\Delta \nu/2} \Delta T_b (\nu^{\prime\prime}) d \nu^{\prime\prime} \\
&=&
\sum_{i=1}^{N({\bar \nu})}{T_s(z_i)-T_{\rm cmb} (z_i) \over {1+z_i}}
e^{-\tau_i ({\bar \nu})} {\Delta \tau_i({\bar \nu})}
\label{simeq}
\end{eqnarray}
where $N({\bar \nu})$ is the number of particles lying along 
the line of sight on the cross section $\Delta S$.
The contribution to the optical depth by a single particle is
\begin{eqnarray}
\nonumber
\Delta \tau_i ({\bar \nu}) 
&=& \int_{\nu-\Delta \nu/2}^{\nu+\Delta \nu/2}
\left({d\tau_i\over d\nu}\right) d\nu^{\prime\prime}\\
&=& {3\lambda_0^3 A_{10} T_\star n_{\rm HI}(z_i)\over 32\pi T_s(z_i) H(z_i)}
\int_{\nu-\Delta \nu/2}^{\nu+\Delta \nu/2} \phi(\nu^{\prime\prime}-\nu_i) d\nu^{\prime\prime}.
\label{taui}
\end{eqnarray}
where $\nu_i$ is the redshifted frequency of the 21-cm line emitted 
from a particle, $i$, and $n_{\rm HI}(z_i)$ is the mean 
density contribution from the particle and is measured by
$n_{\rm HI} = m({\rm HI})_p (\Delta S \Delta d)^{-1}$
where $m({\rm HI})_p$ is the neutral hydrogen mass of the particle
and $\Delta d$ is the spatial depth corresponding to the frequency channel 
width, $\Delta \nu$.
The optical depth to the $i$'th particle is simply a sum of 
$\Delta\tau_j(\nu)$ for intervening particles ($1\le j < i$)
between the observer and the $i$'th particle:
\begin{eqnarray}
\nonumber
\tau_i(\nu) &=& \int_0^{\tau_i} d\tau(\nu)\\
&=&\sum_{j<i} \Delta \tau_j (\nu).
\label{ptau}
\end{eqnarray}
This computation may benefit from sorting and queueing $N({\bar \nu})$ particles
with the distance from the observer.
The discretizations applied in equations (\ref{simeq}) and (\ref{ptau}) are
valid if $\Delta\tau_i$ is sufficiently small \citep{mellema06}.
From equation (\ref{eq-tau}),
we have found that $\Delta\tau \sim 0.03$  at $z=15$ and 
the gas particle remains optically thin in most cases for $10\le z \le 100$
in the dark ages.

The Doppler shift by the pecular velocity is given by \citep{shapiro06}
\begin{equation}
\nu^\prime = \nu {1+\beta \over \sqrt{1-\beta^2}},
\end{equation}
where $\beta \equiv v/c$ and $v$ is the line-of-sight peculiar velocity toward 
the observer.
The last term in the right-hand side of equation (\ref{taui}) was 
reserved for the thermal broadening and of a functional form as
\begin{eqnarray}
\nonumber
&&\int_{\nu-\Delta\nu/2}^{\nu+\Delta\nu/2} \phi(\nu^{\prime\prime}-\nu_i) d\nu^{\prime\prime} =\\
&&{1\over2}\left[{\rm erf}\left({\nu+\Delta\nu/2-\nu_i\over \sqrt{2}\sigma_i(\nu)} \right)
              - {\rm erf}\left( {\nu-\Delta\nu/2-\nu_i\over \sqrt{2}\sigma_i(\nu)} \right)\right],
\end{eqnarray}
where $\sigma_i$ ($ = (\nu_i/c)(k_BT_g/m_{\rm H})^{1/2}$) is 
the one-dimensional (line-of-sight) velocity dispersion.
%In deriving this equation,
%we assumed that each particle has a delta function-like density distribution:
%$\int_0^\infty \mathcal{G}(\nu^\prime) \delta(\nu^\prime -\nu_i) d\nu^\prime=\mathcal{G}(\nu_i)$,
%where $\mathcal{G}(\nu^\prime) \equiv 
%n_{\rm HI}(z^\prime) \phi(\nu^{\prime\prime}-\nu^{\prime}) /T_s(z^\prime) H(z^\prime) $.

\begin{figure}
\center
\includegraphics[scale=0.45]{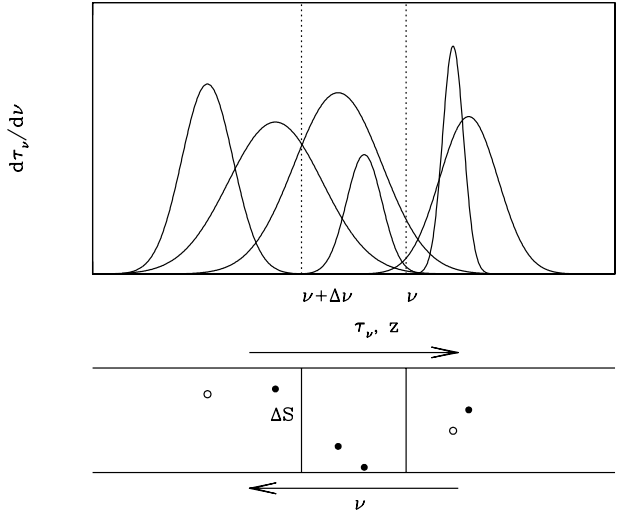}
\caption{ A schematic side view of the line-of-sight optical depth.
An observer is located in the far left side of the figure so that
we can apply the plan parallel approximation.
$\Delta S$ is the two-dimensional cross section 
and $\Delta \nu$ is the frequency width (channel) of the survey element.
The optical depth is growing as moving to right side of the figure while
the redshifted frequency is decreasing. The curves show 
the Gaussian distribution of $d\tau_\nu / d\nu$ ($\propto \phi(\nu)T_s^{-1} H^{-1}(z)$) 
for each gas particle.
}\label{fig-scheme}
\end{figure}

A schematic side view in Figure \ref{fig-scheme} illustrates how to 
measure the brightness temperature in a discrete volume element of 
a depth $\Delta\nu$ and width $\Delta S$.
The observer is assumed to be located at a far left side of the figure 
so that the condition of the plane-parallel approximation can be satisfied.
In the bottom panel, the circles represent simulation particles and central box region
has a  surface area of $\Delta S$ and frequence range $\Delta\nu$.
The line profile broadened by the thermal temperature (here we do not include
the Doppler shift caused by the pecular velocity) are described in the top panel.
Particles which do not substantially contribute to $d\tau_\nu/d\nu$ (and consequently to $T_b$)
between $\nu $ and $\nu+\Delta\nu$ are marked by open circles.
The area of the profile is anticorrelated to the spin temperature $T_s(z_i)$.
The total optical depth is obtained by integrating the Gaussian profiles over 
the frequency range.

It is reasonable to assume that the IGM particles carry an equal amount of
baryonic mass as $m_g = m_p (\Omega_b/\Omega_m)$. 
But this simple relation does not hold any more in the halo region
because the baryonic matter is typically decoupled from the dark matter 
through the hydrodynamic processes.
During the infall it is subject to another force, 
the gas force from neighboring 
baryonic matter due to high densities and temperatures in haloes.
As seen in equation (\ref{eisrho}), the distribution of baryon matter is 
usually different from the NFW profile of dark matter \citep{navarro97}
and, therefore, we should assign a different amount of baryonic mass 
to halo member particles according to modelled distributions of baryonic matter.
First, we measure the matter profile from member particles of
the simulated halo with several radial bins.
Second, we calculate the density ratio of the baryonic matter 
to the simulated matter for a given model of SIS or EIS.
Then, we can compute the amount of baryonic mass to be assigned to 
halo particles for each bin.

\section[]{$N$-Body Simulations \& Halo Findings}
\label{simulation}

\subsection{Simulations \& Halo Findings}
%\begin{deluxetable*}{c|ccccccccccccc}
%\tablecaption{Simulation parameters}
%\tablewidth{0pt}
%\tablehead{
%\colhead{$N_p$}
%&\colhead{$N_m$}
%&\colhead{$L_{\rm box}$}
%&\colhead{$N_{step}$}
%&\colhead{$z_{i}$}
%&\colhead{$z_{f}$}
%&\colhead{$h$}
%&\colhead{$n$}
%&\colhead{$\Omega_m$}
%&\colhead{$\Omega_b$}
%&\colhead{$\Omega_\Lambda$}
%&\colhead{$b$}
%&\colhead{$m_p$}
%&\colhead{$\epsilon$}
%}
%\startdata
%$512^3$ &$512^3$&0.512 & 1188 &300 &15 & 0.719 & 0.96 & 0.258 &0.044&0.742&1.26
%&$71.6 ~h^{-1}{\rm M_\odot}$ &0.1$~h^{-1}${\rm kpc}
%\enddata
%\label{sim}
%\tablecomments{
%Cols. (1) Number of particles
%(2) Number of grids applied to measure the Zel'dovich displacements
%(3) Number of steps
%(4) Initial redshift
%(5) Final redshift
%(6) Hubble parameter
%(7) Spectral index of $P(k)$
%(8) Matter density at $z=0$
%(9) Baryon density at $z=0$
%(10) Dark energy density at $z=0$
%(11) Bias factor
%(12) Particle mass
%(13) Gravitational force resolution
%}
%\end{deluxetable*}
\begin{table*}
\begin{minipage}{160mm}
\caption{Simulation parameters}
\label{sim}
\begin{tabular}{@{}cccccccccccccc@{}}
\hline
$N_p$ & $N_m$ & $L_{\rm box}$& $N_{step}$& $z_{i}$ & $z_{f}$ & $h$ & $n$ & $\Omega_m$ & $\Omega_b$ & $\Omega_\Lambda$ & $b$ & $m_p$ & $\epsilon$ \\
\hline
$512^3$ &$512^3$&0.512 & 1188 &300 &15 & 0.719 & 0.96 & 0.258 &0.044&0.742&1.26
&$71.6 ~h^{-1}{\rm M_\odot}$ &0.1$~h^{-1}${\rm kpc}\\
\hline
\end{tabular}
Cols. (1) Number of particles
(2) Number of grids applied to measure the Zel'dovich displacements
(3) Number of steps
(4) Initial redshift
(5) Final redshift
(6) Hubble parameter
(7) Spectral index of $P(k)$
(8) Matter density at $z=0$
(9) Baryon density at $z=0$
(10) Dark energy density at $z=0$
(11) Bias factor
(12) Particle mass
(13) Gravitational force resolution
\end{minipage}
\end{table*}
We have upgraded the GOTPM \citep{dubinski04} by incorporating
the CAMB Source{\footnote{http://camb.info/sources} \citep{lewis00}
for generating initial power spectrum.
This upgrade is important in this study because the length scales 
of interest are very small ($k\gtrsim 10 ~h{\rm Mpc}^{-1}$)
that the effect of baryons on the matter power spectrum is significant.
For comparison, differences among various power spectra provided
by various method are discussed in Appendix \ref{app:powerspectrum}.
We adopt a cosmological model consistent with the WMAP 5--year cosmology
and set the initial redshift of the simulation $z_i=300$ 
(for the reason of this choice, see the Appendix \ref{app:zeldovich}).
We compute the linear power spectrum of combined matter (CDM + baryonic matter)
at $z=15$ and linearly scale back the power amplitude to $z=300$.
This is because not only does the amplitude of matter power spectrum 
shift with redshift but also the spectral shape changes with time even 
in the early universe. 
Even though we are using the pure $N$-body simulation, we want to obtain 
a simulated power spectrum of the combined matter at $z=15$.  
We call it the core simulation and list several characteristics of the 
simulation in Table \ref{sim}.

\begin{figure}
\center
\includegraphics[scale=0.5]{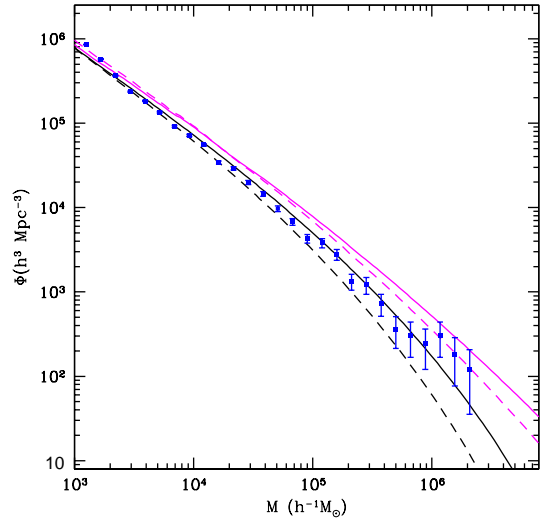}
\caption{
Effect of finite box size on the population of simulated minihaloes at $z=15$.
Each pair of lines are the analytic mass functions of ST ({\it solid})
and PS ({\it dashed line}) 
obtained by integrating over a complete range of the power spectrum ({\it upper two})
and over a range confined to the simulation box ({\it lower two lines}).
Filled boxes are the mass functions of the simulation.
}
\label{fig-mf}
\end{figure}
The friend-of-friend (FoF) halo findings are applied to identify
virialized minihaloes with the simulated particles at $z=15$.
For the linking length, We employ the usually adopted value, $0.2d_{\rm mean}$.
It is interesting to check whether the FoF mass function at high redshifts
is well described by the ST or PS functions 
even on this small scales.
Figure \ref{fig-mf} presents the mass functions of simulated FoF haloes
({\it filled boxes}) and corresponding 
analytic functions such as the ST ({\it solid}) 
and PS ({\it dashed}).
A set of upper two curves is obtained by integrating the power spectrum
over a complete range of wavelength while the other set of two curves is obtained 
by integrating the power over the wavelength confined to the simulation box.
The box-size effect on the halo number density is clearly seen in the figure.
But this effect does not matter in the power spectrum analysis
which is the main topic of this paper.
%Also we show a mass function ({\it open circle}) from a test run with 
%the same simulation characteristics but with a lower starting epoch, $z_i=100$
%to check the effect of the Zeldovich mesh overshooting in the first-order approximation.
%The abundance of the minihaloes is slightly lowered than the reference mass functions
%over mass scales of $M > 10^4 h^{-1}{\rm M_\odot}$.

The simulated mass function is in good agreement with the ST predictions.
This agreement is slightly different from previous results 
of many Lagrangian and Eulerian simulations \citep{wise08,iliev06,reed07,lukic07}
in which authors argued that they have detected 50\% underpopulations of haloes 
compared to the ST predictions at high redshifts.

\section[]{Temperature Maps \& Power Spectrum}
\label{map}

\subsection{Brightness Temperature Maps}
\label{maps}
As described in previous sections, 
we have estimated the kinetic temperature and
local baryonic density at the positions of halo and IGM 
particles, and have allocated these hydrodynamic quantities to $N$-body particles 
so we can treat them as gas particles.
Using these pseudo-gas particles, we run several semianalytic simulations 
targeted for quantifying the effect of semianalytic parameters.
Among these parameters are the switches to turn on or turn off 
the signal sources such as minihaloes and IGM, and 
switches to add the Doppler or thermal distortions on the map. 
Also we measure the IGM temperature
by chosing one of the adiabatic and isothermal processes.
Table \ref{sim} summarizes the semianalytic models we have used in this paper.
The naming convetions are as follows: we use upper cases H or B if the halo or IGM
particles are included in generating the map, respectively.
And the trailing lower script denotes the temperature model of the IGM.
%The trailing upper script $\chi$ means that
%the model uniformly applies the value of IGM ionization fraction even to the halo region.
%This is to contrast the ionization effects on halo signals.
Sometimes, we use trailing marks as (t), (p), or (tp) to denote that 
those models include
thermal broadening, the peculiar-velocity distortion, or both of them, respectively.
We call $\rm HB_{ad}$ a reference model and most of the comparisons are made 
to this model.

\begin{table}
\caption{Semianalytic models}
\begin{minipage}{90mm}
\label{semi}
\begin{tabular}{@{}c cccccc}
\hline
  &  \multicolumn{2}{c}{Signal}
 & \multicolumn{2}{c}{Temperature model} & \multicolumn{2}{c}{Density model}\\
 \cline{2-3} \cline{4-5} \cline{6-7} \\
 Name & {Halo}   & IGM    &
 {Halo}    & {IGM} &
 {Halo}    & {IGM} \\
\hline
$\rm H_{ad}$ & yes & no & EIS\footnote{EIS halo model for temperature} & adiabatic & EIS
\footnote{EIS halo model for density} & \\
$\rm H^{sis}_{ad}$ & yes & no & $\rm SIS$\footnote{SIS halo model} & adiabatic & SIS &- \\
$\rm HB_{ad}$\footnote{the reference model} & yes & yes & EIS & adiabatic & EIS & $W_4$ \\
$\rm B_{ad}$ & no & yes & - & adiabatic & - & $W_4$ \\
$\rm H_{20}$ & yes & no & EIS & 20 K  & EIS & - \\ 
$\rm H_{100}$ & yes & no & EIS & 100 K  & EIS & -\\
$\rm H_{1000}$ & yes & no & EIS & 1000 K  & EIS & - \\
$\rm B_{20}$ & no & yes & - & 20 K & -  & $W_4$ \\
$\rm B_{100}$ & no & yes & - & 100 K & -  & $W_4$ \\
$\rm B_{1000}$ & no & yes & - & 1000 K & -  & $W_4$ \\
$\rm HB_{20}$ & yes & yes & EIS & 20 K & EIS & $W_4$ \\
$\rm HB_{100}$ & yes & yes & EIS & 100 K & EIS & $W_4$ \\
$\rm HB_{1000}$ & yes & yes & EIS & 1000 K & EIS & $W_4$ \\
$\rm \left<HB\right>_{1000}$ & yes & yes & 1000K \footnote{uniform halo temperature fixed to 1000K}
& 1000 K & EIS & $W_4$ \\
\hline
\end{tabular}
Cols. (1) Model name
(2) Halo contribution to signal map
(3) IGM contribution to signal map
(4) Halo temperature model
(5) IGM temperature model
(6) Halo density model
(7) IGM density model
\end{minipage}
\end{table}

\begin{figure}
\center
\includegraphics[scale=0.23]{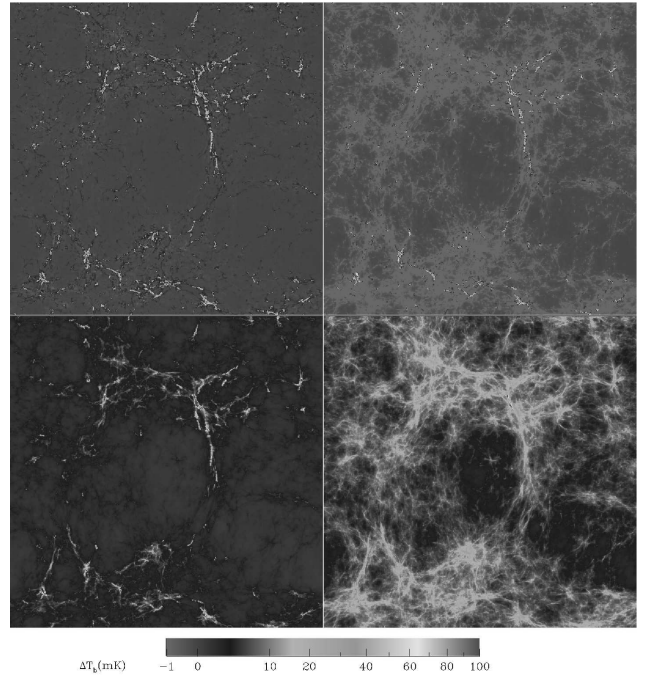}
\caption{
Temperature maps projected along the line of sight in a cubic box of a side length 
$512 ~h^{-1}{\rm kpc}$ in four representative EIS models
(clockwise from upper-left panel, 
$\rm HB_{ad}$, $\rm HB_{20}$, $\rm HB_{1000}$, and $\rm HB_{100}$).
Top-left panel shows temperature fluctuations of haloes and IGM using adiabatic backgrounds.
Other panels show the differential temperature maps obtained by fixing the 
temperature of IGM to $T_{\rm IGM}=20$ K ($\rm HB_{20}$, {\it top-right}),
$T_{\rm IGM}=100$ K ($\rm HB_{100}$, {\it bottom-left}), and
$T_{\rm IGM}=1000$ K ($\rm HB_{1000}$, {\it bottom-right}).
Observed temperatures are measured by averaging brightness temperatures along the line of sight
over $0 < L <512 ~h^{-1}{\rm kpc}$  which corresponds to $88.7402\pm0.0154$ MHz
at $z=15$.
}\label{mosaic}
\end{figure}
Two-dimensional projected temperature maps are shown in Figure \ref{mosaic}
for the reference model ($\rm HB_{ad}$, {\it top-left}) 
and three isothermal models investigated by \citet{furlanetto06b}:
the EIS halo models with isothermal IGM of $T_g=20$ K 
($\rm HB_{20}$, {\it top-right}),
$T_g=100$ K ($\rm HB_{100}$, {\it bottom-left}), and
$T_g=1000$ K ($\rm HB_{1000}$, {\it bottom-right}).
In the $\rm HB_{ad}$ model the average IGM is colder than the CMB 
by about 0.7 mK and, moreover, part of IGM surrounding the overdense filamnetary structures 
is colder than average IGM as noted by \citet{shapiro06}.
If the isothermal IGM is at $T_g=20$K, the observed IGM temperature
is substantially lower than the background CMB temperature compared 
to the adiabatic case (see Fig. 1).
But the IGM brightness temperature becomes higher 
once its temperature is higher than CMB temperature
as can be seen in the bottom-left panel.
As the background IGM temperature is raised, 
the halo signal is getting weaker and haloes become 
less visible. At $T_g=100$K, most of field minihaloes disappear
and only massive minihaloes in crowded regions survive the hot IGM. 
According to the entropy-floor model,
the higher background entropy makes a halo have a bigger but less dense core.
This explains the weaker minihalo signals in the hotter IGM.
At $T_g=1000$K, the diffuse IGM is the dominant source of the observed signals 
showing hot complex structures around dense regions.
Most of the hot signals ($\Delta T_b > 50$ mK) in the model of $T_g=1000$K 
are mainly coming from the hot IGM gas.

\begin{figure}
\center
\includegraphics[scale=0.23]{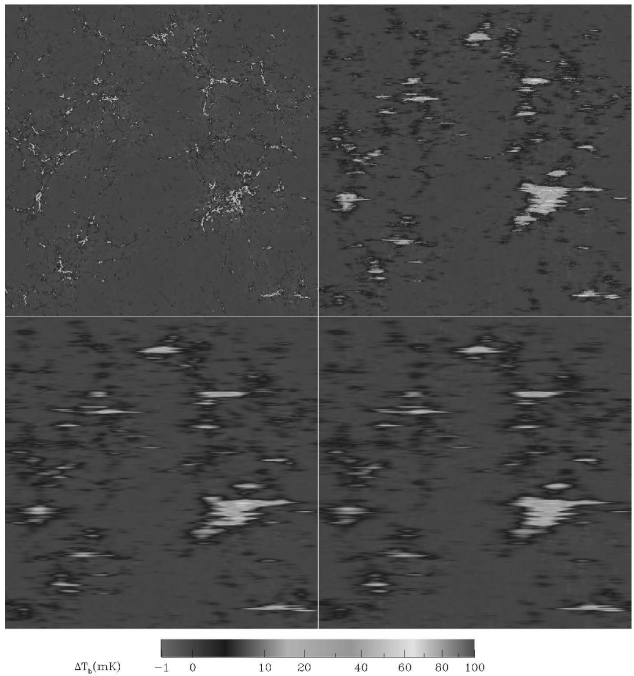}
\caption{
Spectral distribution of brightness temperature 
for the $\rm HB_{ad}$ model ignoring the effects of both the thermal broadening
and peculiar shift ({\it upper-left}), and
considering only the peculiar shift (${\rm HB_{ad}(p)}$, {\it upper-right}),
only the thermal broadening (${\rm HB_{ad}(t)}$, {\it lower-left}),
and both of them (${\rm HB_{ad}(tp)}$, {\it lower-right}).
The $x$ axis of the figure is along the line of sight with a frequency range, $88.7402\pm0.0154$ MHz.
}\label{mosaic2}
\end{figure}
Now, we show the effects of Doppler shift (or peculiar-velocity distortion) 
and/or thermal broadening on the 
redshifted 21-cm map in Figure \ref{mosaic2}.
The $x$ axis of the figure is the line of sight so the observer
is assumed to be far left side of the figure.
The peculiar and thermal distortions make haloes spread along the line of sight and
their effects are especially significant in halo regions.
In the distorted map, the total observed flux in the whole simulation 
box is substantially increased.
This is because the amount of absorption is reduced in the distorted field
as the heavily-obscured emission source appears to be dispersed 
into the less optically thick region in the frequency space.
%the number of 21-cm photons absorbed by the ambient gas is proportional to 
%the variance of $\tau(\nu_i)$ if $\tau(\nu_i)\ll 1$.
%Therefore, the total flux grows as the variance of 
%line-of-sight $\tau$ is decreased. 
%It means that, if the signal flux is spread over the frequency 
%(or the variance of $\tau(\nu_i)$ becomes smaller), photons are less
%absorbed by the intervening neutral hydrogen.
The Doppler shift makes the map noisier than the thermal broadening
in halo regions.
Compared to minihaloes, the IGM experiences less distortions due to the lower temperatures
and smaller peculiar velocities.

\subsection{Effects of Approximation on Power Spectrum}
To justify our semianalytic approach, 
it is crucial to compare our results with the well-known analytic solutions 
or with numerical findings given in other papers.
And one of the most powerful comparisons is using the power spectrum analysis.
We have measured the three-dimensional power spectrum on the signal maps 
and have compared them with those given by 
\citet{furlanetto06b} who measured the signal power spectrum 
of the minihaloes based on the PS function 
under the assumptions of $T_s \gg T_{\rm cmb}$ and $\tau \ll 1$.
For proper comparisions,
we take the same approximations to equations (\ref{simeq}) and (\ref{taui}).
It is worth noting that
the condition of $T_s \gg T_{\rm cmb}$ also satisfies that $\tau \ll 1$.
In this comparison, the switches of peculiar-velocity 
and thermal-broadening are turned off.
\begin{figure}
\center
\includegraphics[scale=0.75]{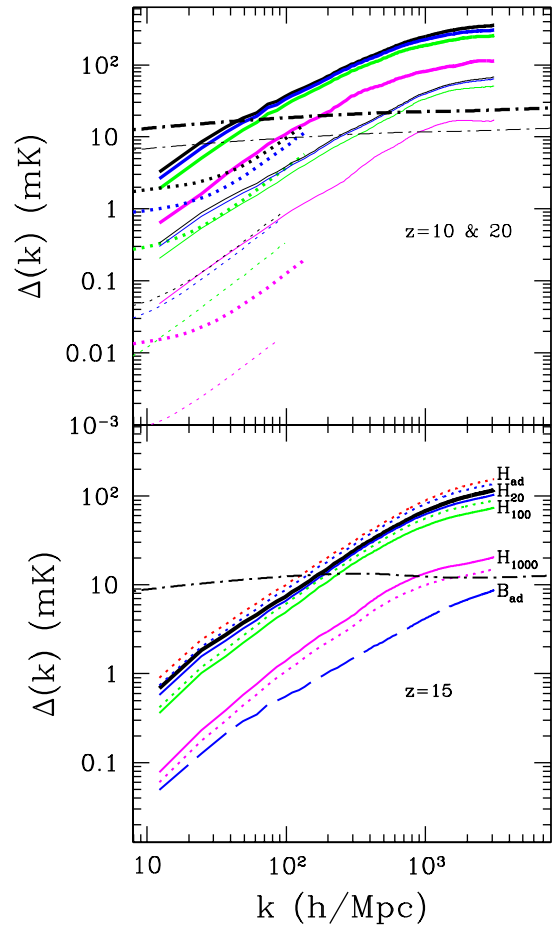}
\caption{
({\it top}):
The halo signal powers at $z=10$ ({\it thick}) and 20 ({\it thin}) 
for the IGM temperature models: adiabatic, $T_g=20$K,
$T_g=100$K, and $T_g=1000$K, from the top most curves of the same styles.
The dotted curves show the predictions of the \citet{furlanetto06b}
while the solid curves are the measurements in this study.
Also the two dot-dashed curves are the predictions
measured using the linear matter power spectrum and the approximation, ($T_s\gg 0$),
at $z=10$ and 20.
({\it bottom}):
Effect of the uniform spin temperature on halo-only powers at $z=15$.
The dotted curves are power spectra obtained under the assumptions of
a uniform spin temperature of a minihalo.
The spin temperature is evaluated at 
the mean overdensity ($\bar{\rho_h} = f_g v_{178}$; \citealt{furlanetto06b}) 
while solid curves are obtained by measuring the spin temperature
for each member particle. 
From the top most in each set of curves we show 
the simulated power spectra of $\rm H_{ad}$, $\rm H_{20}$, $\rm H_{100}$,
and $\rm H_{1000}$ models, respectively.
We plot the $\rm H_{ad}$ model distribution with a thick curve.
The dot-dashed curve is the linear power prediction measured under
the approximations, $T_s \gg T_{\rm cmb}$ and $\tau \ll 1$.
We also plot the background IGM level in long-dashed curve.
}
\label{TsInf}
\end{figure}
To obtain the nonlinear power spectrum of minihaloes, 
\citet{furlanetto06b} have regarded the Fourier transform 
of halo density profile (NFW) as the one-halo term of 
the power spectrum and have assigned to each halo a 
uniform spin temperature measured at the mean gas 
density ($f_g v_{178}$) of the halo.
In their analysis, they used the PS function to populate 
minihaloes in their analysis.
In the top panel of Figure \ref{TsInf}, the dotted curves 
show the signal power of minihaloes reported by \citet{furlanetto06b} 
at $z=10$ ({\it thick}) and $z=20$ ({\it thin}) while two 
sets of solid curves with the same line thickness for the 
same redshift are computed from the numerical simulation in 
this study. 
To get simulated halo data at $z=10$ and 20, we have run an 
auxillary simulation with the same setting as the core simulation
except the initial linear matter power spectrum directly measured 
at $z=300$.
To each redshift data we employ four models: from the topmost curve,
the adiabatic ($\rm H_{ad}$),  $T_g=20$K ($\rm H_{20}$), 
$T_g=100$K ($\rm H_{100}$), and $T_g=1000$K ($\rm H_{1000}$) models.
As can be seen, there are significant deviations between the 
two methods. \citet{furlanetto06b} underestimates the power 
by a few factors at $z=10$ and by an order of magnitude at $z=20$
with a bigger difference in a higher isothermal IGM model.
However, this difference of the signal power agrees with the fact that
the PS predicts less haloes than the ST on massive scales and
more massive haloes have higher power amplitudes.
This discrepancy becomes larger at a higher redshift or in a model
of higher background entropy because halo signals come from more biased 
objects.
The dot-dashed curve shows the linear prediction for the 
matter field of an infinite spin temperature.
In bottom panel, four dotted lines show the power spectra 
of halo signals measured in our simulations at $z=15$
assuming the uniform spin temperature (from the topmost curve, 
$\rm H^\prime_{ad}$, $\rm H^\prime_{20}$,
$\rm H^\prime_{100}$, and $\rm H^\prime_{1000}$)
while a set of solid lines shows the corresponding
power obtained by computing spin temperatures for each 
simulation particles.
%Our measurements ({\it dotted}) are roughly comparable 
%to their's with respect to amplitude and slope 
%except the model of $\rm H^\prime_{1000}$ which shows a considerably larger amplitude than 
%their expectations.
%From the top panel of the figure,
%we come to know that \citet{furlanetto06b} significantly
%underestimate the halo signal powers compared to our measurements
%mainly due to the different number of haloes applied in the analysis.
%The PS function applied by them
%significantly underestimates the number of minihaloes 
%compared to the ST function
%which well describes the simulated minihaloes (see Fig. \ref{fig-mf}).
%From this figure we learn that the uniform spin temperature 
%overproduces the power on most scales indicating that 
%the spin temperature measured at the average density of a halo
%may not be a proper representative for measuring the mean brightness 
%temperature of the halo.

%%%%%%%%%%%%%%%%%%%%%%%%%%%%%
\begin{figure}
\center
\includegraphics[scale=0.45]{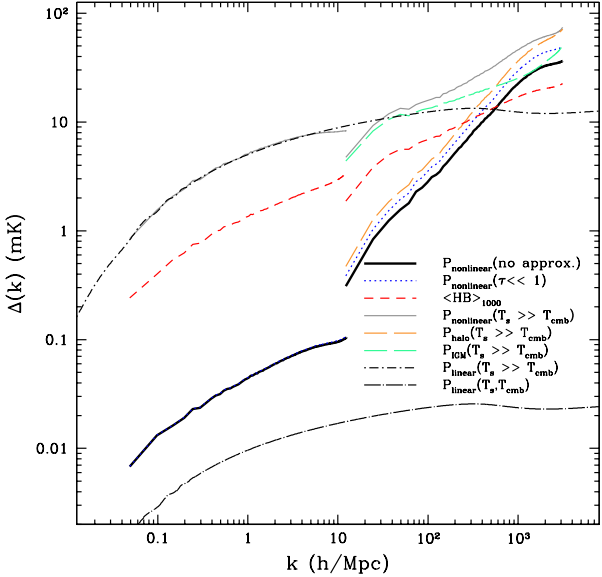}
\caption{
Three dimensional power spectra of the 21-cm signals at $z=15$.
The linear power spectrum for the infinite spin temperature 
($T_s \gg T_{\rm cmb}$) and $\tau \ll 1$ is plotted with a dot-short-dashed curve.
Also, for comparison, the power spectrum measured by setting 
the spin temperature fixed to the mean background value
is shown in the dot-long-dashed curve.
The simulated nonlinear power spectra with infinite spin temperature
are shown in solid gray lines. 
The dotted curves are obtained by assuming the negligible optical depth 
($\tau \ll 1$).  
The thick solid black curves show the power spectra computed without applying 
any approximation to the measurement.
Two long-dashed curves show the halo ({\it thick}) and IGM ({\it thin}) power 
spectra for $T_s\gg T_{\rm cmb}$.
We add the power spectra of the $\rm \left<HB\right>_{1000}$ for comparison
in short dashed curves.
The left set of four simulated power spectra are obtained from another 
simulation of a box size $L_{\rm box}=128 ~h^{-1}{\rm Mpc}$.
}
\label{fig-pkTb}
\end{figure}
Now, our interests are shifted to the effect of the widely used assumptions,
$\tau \ll 1$ and $T_s\gg T_{\rm cmb}$, on the signal power.
Note that the condition, $T_s\gg T_{\rm cmb}$, sufficiently satisfies $\tau \ll 1$
at $z=15$.
We simplify the situation by ignoring the halo model and measure the
density and temperature at the position of every particle 
by the $W_4$ and adiabatic assumption, respectively.
To cover a wider wavelength scale ($k\lesssim 10 h/{\rm Mpc}$), we run another simulation
in a bigger box of a side length, $L_{\rm box}= 128~h^{-1}$Mpc, starting from $z_i=80$.
Figure \ref{fig-pkTb} gives the resulting power spectrum of 21-cm signals 
computed by turning on or off the approximation, $\tau\ll 1$, and
in various temperature models.
The two kinds of dot-dashed curves are obtained from the linear matter power spectrum
but with different approximations:
the dot-short-dashed curve is directly evaluated 
by adopting the approximations that $T_s \gg T_{\rm cmb}$ and $\tau \ll 1$
while the dot-long-dashed curve is measured by
fixing the spin temperature to the mean background value
($\left<T_s^{bg}(15)\right>$).
So we may expect the true adiabatic power spectrum
should lie within these two boundaries in the linear regime.
However, the nonlinear clustering distorts the power spectrum by 
increasing the small scale powers significantly. 
The solid gray curves (${\rm P_{nonlinear}}(T_s\gg T_{\rm cmb})$)
are the simulated nonlinear power for the limits, $T_s \gg T_{\rm cmb}$,
and $\tau \ll 1$. Here we are able to clearly observe the nonlinear clusterings 
on the small scales $k\ge 30~h{\rm Mpc^{-1}}$.
Around $k= 20~ h{\rm Mpc^{-1}}$, the small drop of simulated nonlinear power 
is nothing but the cosmic variance of the simulation.
We call it a baseline model if no approximation is made to the
spin temperature and optical depth.
The thick solid black curve is the power spectrum in the baseline model.
On smaller scales ($k\ge 10 ~h{\rm Mpc^{-1}}$) the slope is much steeper
than the linear (${\rm P_{linear}}(T_s\gg T_{\rm cmb})$) and simulated 
(${\rm P_{nonlinear}}(T_s\gg T_{\rm cmb})$) models of the infinite spin temperature.
However, the power shapes are similar to each other on larger scales.
The dotted curve is plotted to show how significantly the signal power 
deviate from the baseline model by the assumption of $\tau \ll 1$.
The assumption of negligible optical thickness results in a rise to the small-scale power 
at $k\gtrsim 10 ~h{\rm Mpc^{-1}}$ but leaves no effect on the larger-scale signals.
This scale-dependent deviation in the power spectrum may come from the 
scale-dependent fluctutations of optical thickness.  
On larger scales ($k < 10~h {\rm Mpc}^{-1}$), the density fluctuation 
does not develope so much that the optical depth is generally negligible
while small-scale structures are well developed and may have 
substantially higher optical thickness.

It is valuable to note a change of the amplitude 
and slope of power spectrum when an assumption on the spin temperature is applied.
For comparison, we overplot the power spectrum of $\rm \left<HB\right>_{1000}$ model
({\it thick-dashed}). This model is added to check whether 
the gas temperature of $T_g=1000$K would be enough to
make the spin temperature sufficiently high to satisfy the 
infinity approximation ($T_s\gg T_{\rm cmb}$).
The model predicts a much lower amplitude 
by a factor of about five than the $\rm P_{nonlinear} (T_s \gg T_{\rm cmb})$ model
but with the same shape.
This amplitude difference may be caused by the insufficient gas density 
which finds it hard to efficiently pump up the spin temperature at $z=15$.
Even though the gas temperature reaches $T_g=1000$K, the spin temperature 
of a region of $\delta_\rho=0$ (10) could only rise to $T_s \sim 50$K (100K).
From the matter power spectra of the halo and IGM, we measure 
the signal power spectra under the approximation of $T_s \gg T_{\rm cmb}$.
In the figure, the halo power ({\it thick long-dashed}) has a steep slope 
and crosses the IGM power ({\it thin long-dashed}) 
at $k \simeq 6\times10^2~ h{\rm Mpc}^{-1}$.
We do not draw the halo power in the bigger-box simulation because no virialized haloes
($M_h \ge 3.4 \times 10^{10} ~ h^{-1}{\rm M_\odot}$) are identified due to a low mass resolution.
%The relation between the components of power spectrum is written as
%\begin{equation}
%P_{\rm all} = \left( 1 - \mathcal{T}\right)^2 P_{\delta_m\delta_m}
%+\left(2\mathcal{T}\right) P_{\delta_m \delta_{Ts}}
%+\left(\mathcal{T}^2\right) P_{\delta_{T_s} \delta_{Ts}},
%\end{equation}
%where $\mathcal{T} \equiv {T_{\rm cmb}/T_{s0}}$
%and $T_{s0}$ is the mean spin temperature.
%If $T_{s0} \rightarrow \infty$, we get $P_{\rm all} = P_{\delta_m\delta_m}$.
%This smaller amplitude comes from the fact that $(1-T_{\rm cmb}/T_s) < 1$
%for all cases.

\subsection{Power Spectrum of haloes and IGM}

\begin{figure}
\center
\includegraphics[scale=0.45]{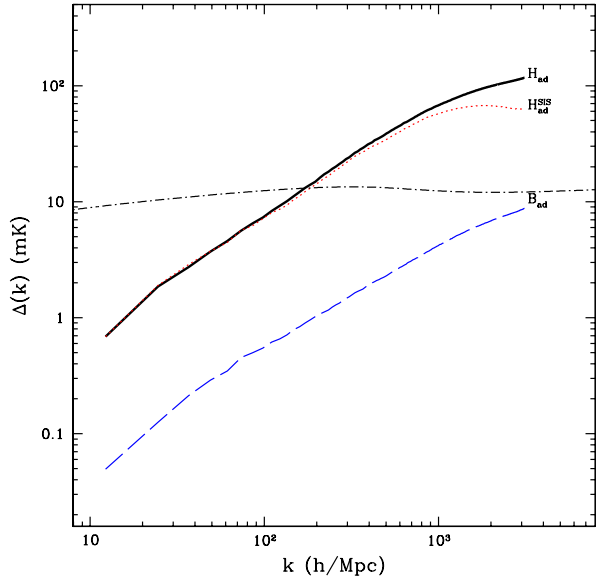}
\caption{ 
Halo power spectra in SIS and EIS models.
The solid line shows the halo power of the entropy model ($\rm H_{ad}$), and
the dotted line is for the SIS haloes (${\rm H_{ad}^{SIS}}$).
For comparison, the power spectrum of adiabatic IGM ($\rm B_{ad}$) 
is shown in the dashed curve.
The dot-dashed curve is the linear prediction for $T_s \gg T_{\rm cmb}$ and $\tau \ll 1$.
}
\label{pk}
\end{figure}
We want to highlight the effect of the entropy on the power spectrum of 
halo signals.
The background IGM signals are excluded in order to isolated the role 
of entropy in the halo signals.
Figure \ref{pk} shows the signal power spectra of haloes in the two models 
(SIS in {\it dotted} and EIS in {\it solid} curves).
The model difference is a function of the scale and increases with wavenumber, $k$.
The EIS model shows a slightly higher amplitude than the SIS model 
because of the higher core temperature and lower gas density
which produces higher spin temperature and lower optical thickness in haloes,
which leads to higher signal fluctuations (see Eq. \ref{eq-htb}).
The background power of adiabatic IGM ({\it dashed}) has an amplitude 
an order of magnitude lower than all the available halo models but 
has a similar power slope.

\begin{figure}
\center
\includegraphics[scale=0.45]{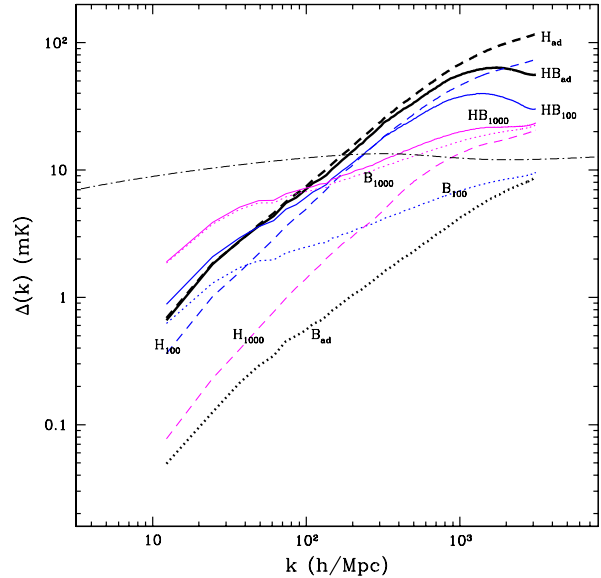}
\caption{
The contrast of minihalo signals in the background IGM emission.
The power spectra of redshifted 21-cm signals of minihaloes ({\it dashed})
and IGM ({\it dotted})
are shown against that of combined signals ({\it solid curves})
for ${\rm HB_{1000}}$, and ${\rm HB_{ad}}$ models.
Each line is tagged with the applied model name.
}
\label{pkiso}
\end{figure}
It is important to note the characteristic scale below which halo signals 
begin to dominate the diffuse backgrounds.
Figure \ref{pkiso} shows the power spectra of the halo ({\it dashed}),
IGM ({\it dotted}), and both of them (halo+GIM {\it solid}) 
in three different halo models.  In the ${\rm HB_{1000}}$, the power spectrum 
is dominated by the IGM and halo signals are completely buried in it.
Therefore, if the IGM was preheated to $T_g=1000$K,
we can not observe the minihaloes in the redshifted 21-cm observations.
Meanwhile, the $\rm HB_{100}$ model has
a dominant IGM power on larger scales ($k<50~ h{\rm Mpc}^{-1}$).
But, on the other hand, the halo signal is stronger on the smaller scale.
In the ${\rm HB_{ad}}$, the halo power is much higher than the IGM
showing the strongest power compared to isothermal models.
Preheating before minihalo formation may be an inportant factor
to expect whether minihaloes can be observed in the dark ages.
A higher temperature of IGM more strongly suppresses the halo signals,
leaving hotter and bigger cores of protogalaxies in minihaloes.
The power amplitude of the isothermal IGM is an increasing function of the gas temperature
and the power spectrum is steeper than the adiabatic IGM.
Therefore, the slope of combined power spectrum depends on the IGM temperature 
at the epoch of minihalo formation.
A steeper slope of the power prefers a lower IGM temperature and 
by observing the slope of the power at these scales we will know the preheating history of the IGM.
The adiabatic IGM has a more steeper slope ($n_\Delta \sim1 $) while the IGM preheated to $T_g=1000$K
has a slope of $n_\Delta \sim 0.5$ over $10 \le k \le 1000 ~ h{\rm Mpc^{-1}}$.

\subsection{Effects of Thermal Broadening and Doppler Shifting}
As easily seen on the temperature map in Figure \ref{mosaic2}, 
the thermal and Doppler broadening
are significantly stretching hot halo signals along the line of sight.
Rich structures in the crowded regions of haloes are significantly smoothed out
and a large fraction of small minihaloes in the mean fields are severly
buried in the IGM due to the flattening.
In Figure \ref{pk2} we show changes of the power spectrum 
from the reference model owing to the peculiar velocity 
({\it short-dashed}) and thermal broadening ({\it dotted}). 
The peculiar velocity tends to enhance the large scale powers \citep{kaiser87}
but lowers the small scale powers
as typically observed in the galaxy redshift surveys \citep{heavens98,park94}.
The power spectrum of biased objects
distorted by the exponential or Gaussian velocity dispersions
is well described by \citep{park94,cole95}
\begin{equation}
P(k,\mu) = P^R(k) { \left(1+\beta\mu^2\right)^2 \over \left(1+k^2\sigma_v^2 \mu^2/2\right)^2},
\label{gauss}
\end{equation}
or 
\begin{equation}
P(k,\mu) = P^R(k)  \left(1+\beta\mu^2\right)^2 \exp{\left(-k^2\sigma^{\prime2}_v\mu^2/2\right)},
\label{exp}
\end{equation}
respectively.
Here $P^R(k)$ is the real-space power spectrum, $\beta\simeq \Omega^{0.6}(z)/b_{21}$,
$b_{21}$ is the bias factor for 21-cm signals,
$\sigma_v$ is the velocity dispersion, and $\mu$ is the directional cosine
of the wave vector along the line of sight.
The former equation has been known for better description of the distribution 
of simulated peculiar velocity \citep{park94} while the latter one is better 
for the Gaussian velocity dispersion like the thermal broadening.
By simply scaling the large scale power we obtain $b_{21} = 0.8$ 
since $\Omega(z)\simeq1$ at the redshift of interest.
Also the one-dimensional velocity dispersions 
of the simulated particles give that
$\sigma_v = 2.48(1+z){\rm km~ s^{-1}H^{-1}}(z)$ for the peculiar velocity
and $\sigma^\prime_v = 2.13 (1+z){\rm km~ s^{-1}H^{-1}}(z)$ for the thermal broadening.
Here, the redshift term is multiplied to change the scale from 
the real space to comoving space.
We integrated above equations over $\mu$ and measured the averaged power spectrum.
The thick gray curves in the figure show those predicted power spectra whose
amplitudes are boosted up by $1/b_{21}$ to match for the global amplitudes
of the simulated power.
On the scale of $k \ge 300~ h{\rm Mpc^{-1}}$
the amount of reduced power is much larger than expected for both distortions
and this discrepency between the expectation and measurment may come from the
change of the optical depth as argued in the previous subsection.
\begin{figure}
\center
\includegraphics[scale=0.45]{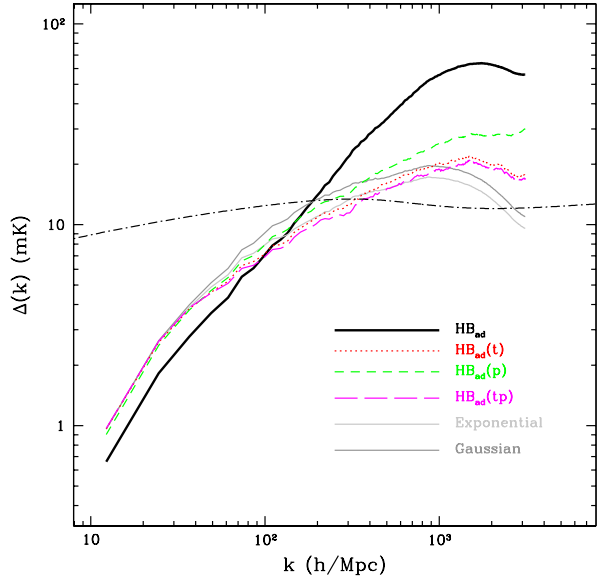}
\caption{Power spectrum distortions caused by the thermal 
broadening and peculiar shift at $z=15$ in the reference model.
The thick solid curve is measured when we do not apply any of the distortions.
The dotted curve is obtained if only the thermal broadening effect is considered and
the effect of the Doppler shift is shown in the short dashed curve.
The long dashed curve is the resulting power spectrum 
when two effects are simultaneously considered.
Two gray lines are the predicted power spectrum. The dark gray and light gray
are predicted from the Gaussian and exponential distribution models, respectively.
%Power spectra of $\rm H_{ad}$ and $\rm HB_{ad}$ are almost overlapped with each other.
}
\label{pk2}
\end{figure}

Now we study the effects of redshift distortions on the halo and IGM fields separately.
In Figure \ref{pkisopk}, the power spectra of the two distorted fields are shown
with the same line types as shown in Figure \ref{pk2}.
To this study, we apply the thermal and peculiar distortions together.
As seen in equations (\ref{gauss}) and (\ref{exp}),
there is nearly no global amplitude increase in the IGM power
because the IGM has a zero bias ($b=1$) and $\Omega(z) \simeq 1$.
But the decrease of IGM power on the small scale depends on the temperature.
In the $\rm HB_{ad}$ and $\rm HB_{100}$ models, the $\Delta(k)$ have a peak around 
$k= 1400 ~h {\rm Mpc}^{-1}$
while $\rm HB_{1000}$ model has a power spectrum peak at $k=80~h {\rm Mpc}^{-1}$.
We learn that haloes are a more dominating factor in the signal power 
than the background IGM on minihalo scales.
Also in the $\rm HB_{100}$ model haloes are more powerful on the scale below
$k\simeq  30 ~h{\rm Mpc}^{-1}$. And the most of the power spectrum in the $\rm HB_{1000}$ 
model comes from the IGM signals.
The distorted IGM power explored in this paper is always smaller than the linear prediction 
estimated by using $T_s\gg T_{\rm cmb}$.
The slope of the power spectrum is considerably flatter than the non-distortion case:
the $\rm HB_{1000}$ model shows a nearly flat power spectrum on $100< k < 1000 ~h{\rm Mpc}^{-1}$
while $\rm HB_{ad}$ model predicts a spectral index of $n_\Delta \simeq 0.5$ on the 
same scale.
\begin{figure}
\center
\includegraphics[scale=0.45]{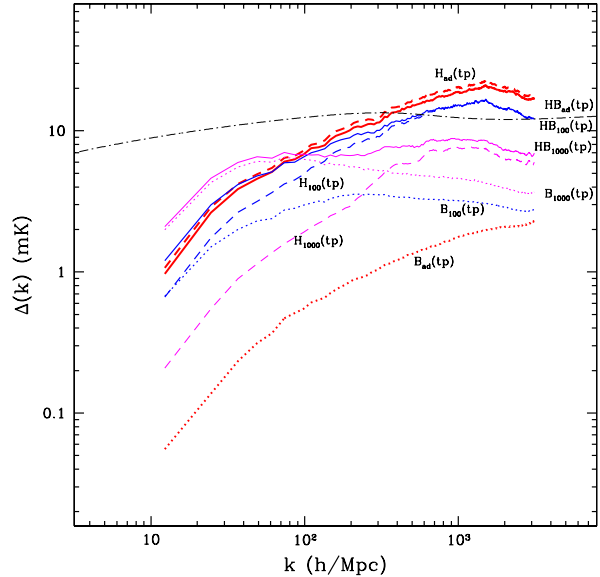}
\caption{
Same as Fig. \ref{pkiso} but including the distortion effects.
Power spectra of $\rm H_{ad}(tp)$ and $\rm HB_{ad}(tp)$ are almost overlapped with each other.
}
\label{pkisopk}
\end{figure}

\section[]{Conclusions \& Discussions}
\label{conclusion}
We have proposed a new semianalytic method to map the hydrogen distributions 
in the dark ages based on the Lagrangian data of the N-body simulation.
One of the most favourable features of the method is that
it adopts a robust way of the optical depth measurement.
Also the entropy-floor model is applied to properly describe the 
temperature and baryonic density in minihaloes.

By analysing the power spectrum of the generated maps, we learn 
that haloes in the entropy-floor model dominate the adiabatic IGM 
in 21-cm signals over the entire scale 
($10 \le k \le 3\times 10^{3} ~h{\rm Mpc}^{-1}$)
available in the simulations.
However, the signal fluctuations of IGM with $T_g=1000$K
significantly overwhelm the minihalo signals over all scales.
And the model with the IGM of a small temperature $T_g=100$K 
predicts that the halo power spectrum is larger than 
the IGM on the scale of $k \gtrsim 50 (30)~h{\rm Mpc}^{-1}$ on 
the distorted (non-distorted) maps.
Because the power spectra of the IGM and halo have different slopes 
to each other, we may observationally determine the preheating temperature 
of the IGM from the power slope on the minihalo scales.
Moreover, the thermal broadening and the peculiar velocity distortions 
make the slope more flattened.

The adiabatic power spectrum of the matter field is on large scales 
five times larger than the linear prediction for a spin temperature
fixed to the background value as $T_s = \left<T_s^{bg}\right>$ at $z=15$.
But on small scales ($k>10 h{\rm Mpc}^{-1}$), the power spectrum 
rises more steeply with $k$ than the linear model and becomes even larger 
than the {\it upper-bound} linear model of $T_s\gg T_{\rm cmb}$
at $k \simeq 600~h{\rm Mpc}^{-1}$.

Test measurements of power spectrum with several approximations 
frequently employed in the literature have shown substantial 
deviations around the true values.
We have found that the worst case is the adoption of the Press 
\& Schechter function for the minihalo number density at early universe.
The Press \& Schechter function underestimates the number of massive 
minihaloes compared to the simulation results which, on the other hand,
is well described by the Sheth \& Tormen function. 
Also the uniform spin temperature produces a considerably larger power 
than the true value.
Either of the peculiar velocity and thermal broadening seriously 
affects the power spectrum making the signal fluctuations globally boosted
while significantly suppressing the small-scale fluctutions.
Although this effect of the redshift distortion on the power spectrum
is similar to that of the galaxy surveys, we found that the analytic 
exponential or Gaussian distributions gives a poor fit to 
the simulated power spectrum on small scales.
One possible explanation for this difference may be the decrease 
of the number of absorbed photons in the distorted field. 
The heavily obscured regions are stronly subject to the distortions
and the photons emitted in the region can be less absorbed by the 
intervening hydrogen atoms because wavelengths of emitted photons 
are shifted to neighboring region where the optical depth would be lower.
As a resut, this effect may suppress the small-scale signals less 
efficiently than expected.

There are two assumptions employed in our numerical method.
First, a gas particle is assumed to have a volume 
not being overlapped with the other particles.
To measure the optical depth along the line of sight,
simulation particles are sorted and queued in order of distance from the observer.
Particles are considered to have their own exclusive box-shaped volumes of 
equal density
and they are stacked along the line of sight with the same cross section.
This approximation allows us to build a simple expression for optical depth 
as shown by equation (\ref{ptau}).
If particles are allowed to have smooth density profile 
and they can overlap with each other,
the equation of optical depth would be complicate and 
the implementation would be difficult.
However, it would not seriously change the conclusions arrived in this study
because the discreteness effect on the particle volume on the optical depth
may be small and only confined to the resolution scale.
Also the method does not consider the Wouthuysen-Field
by the $Ly\alpha$ photons emitted from stars 
or AGNs, which is very important in the reionization epochs. 
In the future extension of the method, we will cover this issue. 

Second, we assume the simulated haloes to be spherical.
In this semianalytic method, we have allocated baryonic mass and temperature to the
$N$-body particles.
To generate protogalaxies in minihaloes, 
we subgrouped member particles with multiple spherical shells of an equal centre.
Because we already know the radial density profile of baryons in the EIS or SIS model,
we can determine how much baryonic mass should be allocated to each bin particle.
However, a problem happens when the distribution of simulation particles
of a halo is aspherical, which is common in most haloes.
In this non-spherical case, the derived shape of the protogalaxy 
built by our method is also following the shape of the simulated minihalo. 
So the spherical assumption in the EIS model is not valid any more. 
Therefore, care must be taken to the fact that the resulting baryonic distribution 
of minihaloes could be aspherical.

\section*{Acknowledgments}

This work is supported by the Korean Research Foundation Grant
KRF-2008-357-C00050 funded by Korean government. 
JHK thanks CITA for the hospitality during his visit.
Simulations and most of subsequent analysis 
were carried out on the Sunnyvale, a linux cluster at CITA.

\appendix
\section[]{Simulation Power Spectrum}
\label{app:powerspectrum}
Comparing the linear power spectra proposed in the literature is an interesting task
and computing an exact power spectrum on small scales is crucial for the study of
formation and evolution of haloes especially in the early universe.
For the CDM power spectrum Eisenstein \& Hu (1998) proposed a simple fitting function
that has been widely used for its simplicity, fast speed, and easy implementation.
Later, the CAMB \citep{lewis00} based on the CMBFAST \citep{seljak96}, 
provides a fast (but not so much fast as EH method)  
and error-controllable package for the power spectrum.
There is a variant of CAMB, the CAMB Source,
which calculates more completely the small-scale transfer function 
considering the baryon sound speed.
Figure \ref{fig-diffpk} shows the differences of 
the CAMB and CAMB Source from the Eisenstein \& Hu's fitting function.
In the left panel shown are the WMAP 5-year power spectrum linearly extrapolated to $z=0$.
There seems no obvious deviation between them below $k\sim 100 ~h {\rm Mpc}^{-1}$. 
But if plotted in the differential power spectrum,
we are able to notice their differences more easily.
The differential power spectrum defined by
$\delta P(k) \equiv (P(k)-P_{\rm EH} (k))/P_{\rm EH}(k)$
are shown in the right panel of the figure. 
The solid curve shows the difference of CAMB Source 
from the EH and the dotted curve is for CAMB.
In this figure, we note that the EH model estimates the matter powers 
within a few percent accuracy on the large scale ($k>10^2 ~ h/{\rm Mpc}$).
And intriguingly, the CAMB slightly underestimates the small-scale power
compared to the CAMB Source and can not properly recover the periodic ripples
either on small scales.
The difference of the power amplitude between the CAMB Source and Eisenstein \& Hu
is, on average, less than 4\% with a maximum difference of 7\% around 
$k\simeq 0.08 ~h{\rm Mpc^{-1}}$ for $10^{-4} \le k \le 0.3 ~h{\rm Mpc^{-1}}$.
But for scales of interest in this study ($10 \le k \le 3\times 10^3 ~h{\rm Mpc^{-1}}$), 
the power spectrum difference
between the CAMBs and the Eisenstein \& Hu's becomes significant with high $k$ value.
This difference is significant when measuring the power spectrum of matter or 
redshifted 21-cm signals. 
So it is very important to use the CAMB Source for generating the initial power spectrum
for the minihalo studies in the early universe.
\begin{figure*}
\center
\includegraphics[scale=0.9]{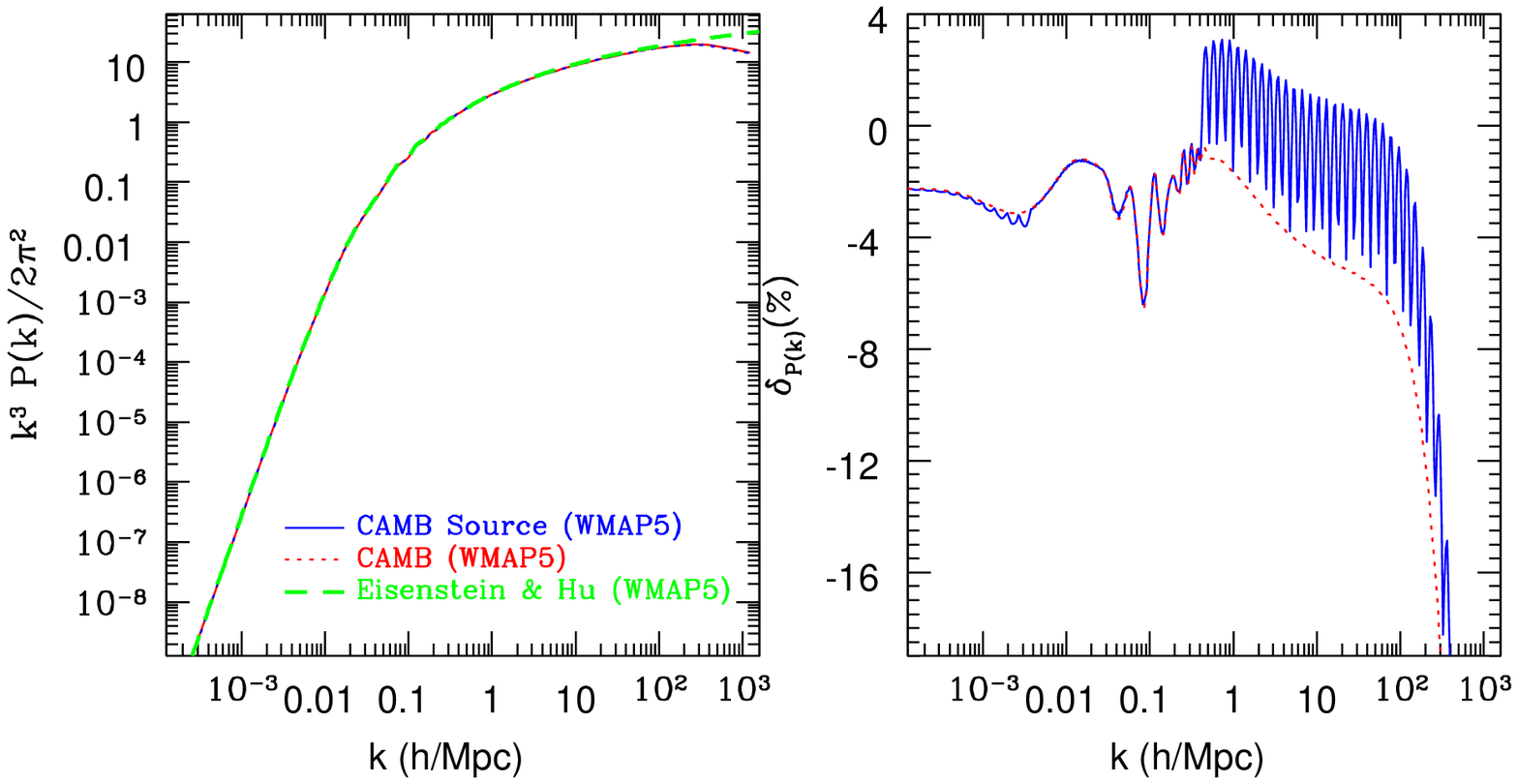}
\caption{
Matter power spectrum at $z=0$.
({\it Left panel}): Power spectra of the Eisenstein \& Hu ({\it dashed}),
CAMB ({\it dotted}), and CAMB Source ({\it solid}) are shown based on the WMAP 5-year 
cosmology.
({\it Right panel}): Deviations from the Eisenstein \& Hu are shown
for the CAMB ({\it dotted}) and CAMB Source ({\it solid curves}).
All the power spectra are generated for the WMAP 5-year cosmology.
}\label{fig-diffpk}
\end{figure*}

\section[]{Setting Initial Redshifts \& Its Effect on the Halo Mass Functions}
\label{app:zeldovich}
\begin{figure*}
\center
\includegraphics[scale=0.9]{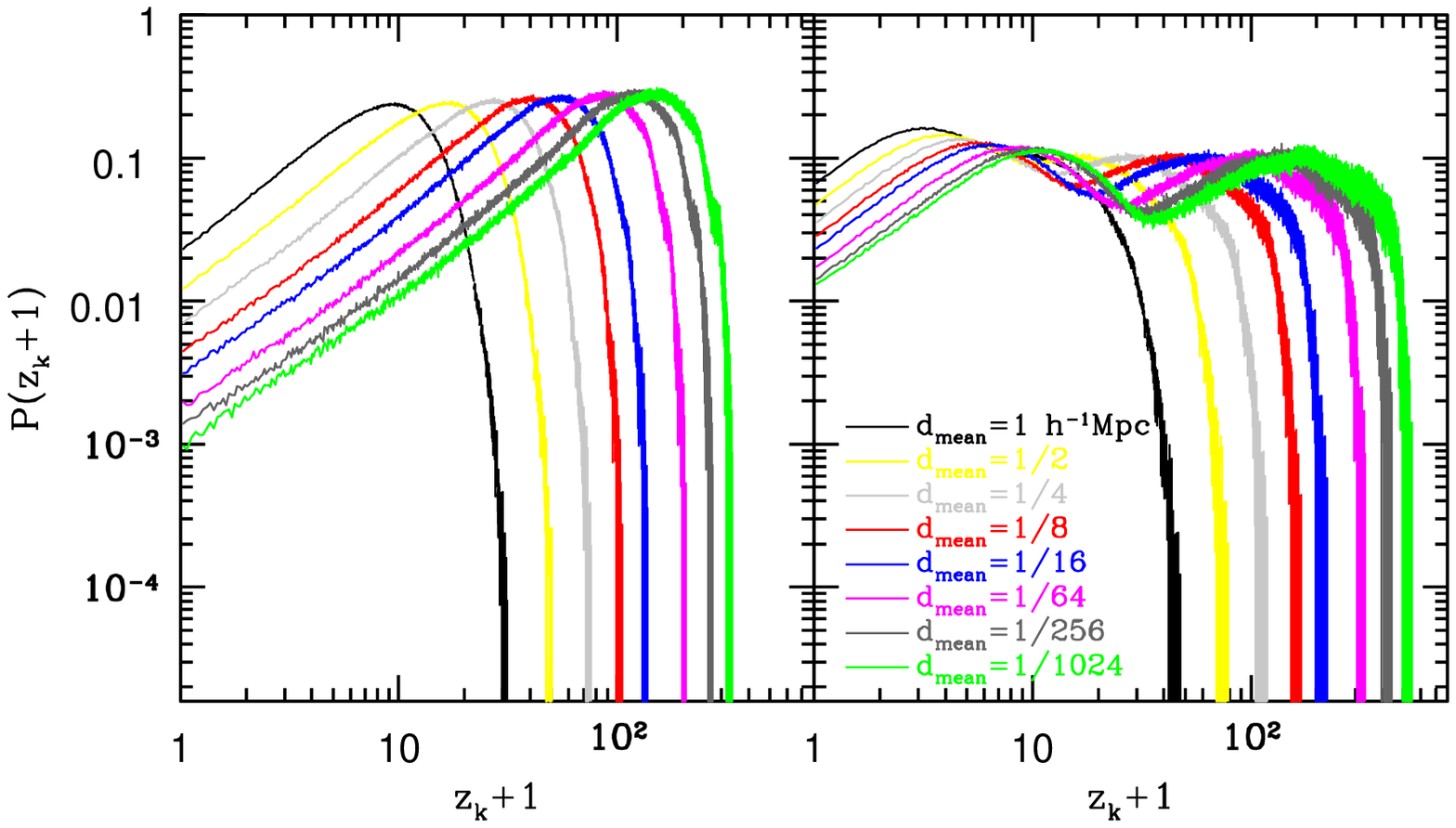}
\caption{
The probability distributions of $z_{pp}$ ({\it right}) and $z_{pm}$ ({\it left}).
The simulation resolution is written in the legend
in terms of the mean particle separation.
}
\label{fig-zel}
\end{figure*}
The GOTPM code adopts a first-order Lagrangian perturbation scheme to generate
initial conditions. It is faster and simpler than the second-order scheme \citep{crocce06}.
But the initial conditions should be generated with much care 
to satisfy assumptions adopted in the scheme.
Like the definition in the appendix of \citet{kim09}, the Zel'dovich redshift ($z_k$) of a particle
is the redshift when the displacement of a particle from its Lagrangian point 
is equal to the mesh spacing either in $x$, $y$, or $z$ direction.
Prior to setting up the initial conditions, particles are located at the mesh points and
linear velocities are assigned to them according to the Zel'dovich approximations
from which the initial displacements are calculated.
Therefore, if the displacements are greater than the mesh size,
they can not reflect the mesh-size fluctuations of the Zel'dovich potentials
in the first-order scheme.
So the initial starting redshift should properly be less than the Zel'dovich redshifts.

There is an alternative to this definition of the Zel'dovich redshift: 
the redshift when the relative separation between neighboring particles is zero
in one of the three dimensions. This type of the Zel'dovich redshift 
would be the proper one if one wants to avoid the situation when two 
adjacent particles overshoot each other in the initial conditions. 
This constraint on the starting redshift would be more stringent than the original one.
We call this the particle-particle Zel'dovich 
redshift ($z_{pp}$) and the former one is renamed the particle-mesh 
Zel'dovich redshift ($z_{pm}$).
In Figure \ref{fig-zel}, we show the distributions of $z_{pm}$ ({\it left}) 
and $z_{pp}$ ({\it right}) 
for simulation resolutions expressed in terms of the mean particle separation ($d_{\rm mean}$).
In the simulations we use $256^3$ particles and the same number of mesh
to measure the initial displacement from the generated Zel'dovich potential.
Therefore, one should note the lack of large-scale power in these ``gauge'' tests
and the underestimation of the Zel'dovich redshifts are expected 
especially for the small $d_{\rm mean}$ cases.
For the $z_{pm}$ distribution, all distributions show an almost same shape
with a single peak around which a power-law rising and a sharp cut off with $z_{pm}$ are shown.
And the $z_{pp}$ distribution has double peaks with drops at higher redshifts
than $z_{pp}$.

\begin{figure*}
\center
\includegraphics[scale=0.9]{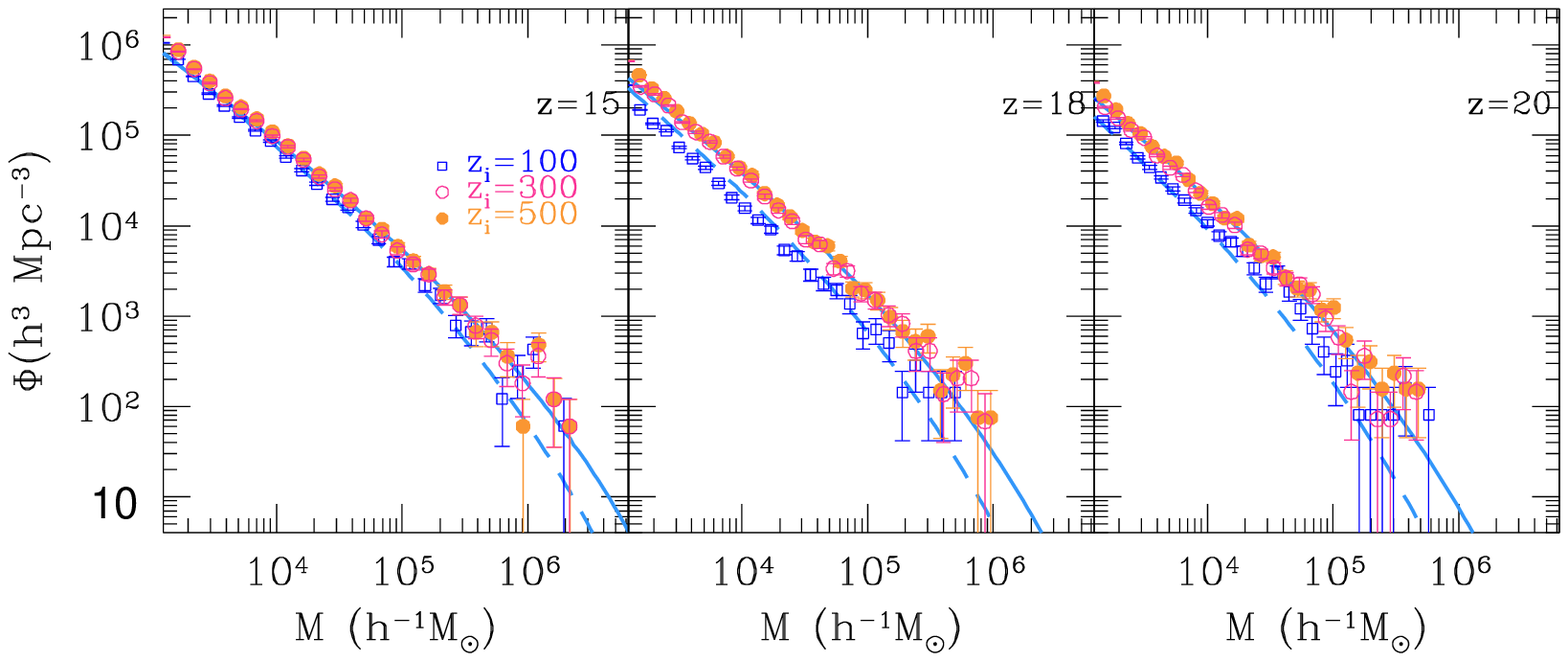}
\caption{
FoF halo mass functions at $z=15$ ({\it left}), $z=18$ ({\it center}), and $z=20$ ({\it right}).
Open boxes show halo abundances from the simulation
with a starting redshift, $z_i=100$ while open circles and filled circles
represent the mass functions for $z_i=300$ and $z_i=500$, respectively.
The solid curve follows the ST function and
the dashed curve marks the PS function
computed with the power spectrum confined to the box of a side length $L_{\rm box}=0.512 ~h^{-1}{\rm Mpc}$.
}
\label{fig-b2}
\end{figure*}
One of the easiest and simplest ways to justify the selection of the initial redshift is 
to measure the abundance of simulated haloes at later epochs.  
Simulations starting with redshifts equal to or higher than
a certain critical epoch should show the same halo mass functions,
and simulations with lower starting redshifts may have deviations 
from the true distribution.
Figure \ref{fig-b2} emphasizes the importance of $z_{pm}$ in determination of the initial redshift.
For simulations with cubic boxes of side length, 
$L_{\rm box}=0.512 ~h^{-1}{\rm Mpc}$ and $512^3$-size mesh,
we select three chracteristic initial redshifts for comparison.
First, $z_i=500$ and $z_i=300$ are chosen from distributions of $z_{pp}$ and $z_{pm}$, respectively.
And then, $z_i=100$ is added to contrast the lower starting redshift against above higher values.
As can be seen in Figure \ref{fig-zel}, most of initial particles at $z_i=100$
are shifted larger than the mesh spacing.
Using these initial settings, we run the simulations down to $z=15$.
The numbers of time steps are determined to satisfy that the maximum displacement of
particles in a step be less than $0.1 d_{\rm mean}$ 
which is also set equal to the force resolution. 
The total time steps from $z_i$ to $z=15$ are set nearly same 
(ranging from 1,063 for $z_i=100$ to 1,214 for $z_i=500$ simulation)
in the three simulations.
It is interesting to note that at $z=18$ and 20 the simulation with $z_i=100$
has underpopulations of haloes ($\sim$ 50\%) compared to other simulations 
while the abundance differences narrows to a few percent level at $z=15$.
There is no obvious difference between the $z_i=300$ and $z_i=500$ simulations
so we conclude that it would be better to use the $z_{pm}$ distribution for
the starting redshift of simulations.

\end{document}